\documentclass[twocolumn,english,aps,superscriptaddress]{revtex4}
\usepackage[T1]{fontenc}
\usepackage[latin9]{inputenc}
\usepackage{color}
\usepackage{babel}
\usepackage{amsmath}
\usepackage{amssymb}
\usepackage{graphicx}
\usepackage{esint}
\usepackage[unicode=true,pdfusetitle,
 bookmarks=false,
 breaklinks=false,pdfborder={0 0 0},backref=false,colorlinks=true]
 {hyperref}

\makeatletter

\newcommand{\lyxdot}{.}

\@ifundefined{textcolor}{}
{%
 \definecolor{BLACK}{gray}{0}
 \definecolor{WHITE}{gray}{1}
 \definecolor{RED}{rgb}{1,0,0}
 \definecolor{GREEN}{rgb}{0,1,0}
 \definecolor{BLUE}{rgb}{0,0,1}
 \definecolor{CYAN}{cmyk}{1,0,0,0}
 \definecolor{MAGENTA}{cmyk}{0,1,0,0}
 \definecolor{YELLOW}{cmyk}{0,0,1,0}
}

\@ifundefined{showcaptionsetup}{}{%
 \PassOptionsToPackage{caption=false}{subfig}}
\usepackage[caption=false]{subfig}
\makeatother

\begin{document}

\title{Near field radiative thermal transfer between nano-structured periodic
materials}

\author{Hamidreza Chalabi}

\email{chalabi@stanford.edu}

\affiliation{Geballe Laboratory for Advanced Materials, Stanford University, Stanford,
California 94305, USA}

\author{Erez Hasman}

\email{mehasman@technion.ac.il}

\affiliation{Micro and Nanooptics Laboratory, Faculty of Mechanical Engineering,
and Russel Berrie Nanotechnology Institute, Technion-Israel Institute
of Technology, Haifa 32000, Israel}

\author{Mark L. Brongersma}

\email{brongersma@stanford.edu}

\affiliation{Geballe Laboratory for Advanced Materials, Stanford University, Stanford,
California 94305, USA}

\date{12/28/13}
\begin{abstract}
This paper provides a method based on rigorous coupled wave analysis
for the calculation of the radiative thermal capacitance between a
layer that is patterned with arbitrary, periodically repeating features
and a planar one. This method is applied to study binary gratings
and arrays of beams with a rectangular cross section. The effects
of the structure size and spacing on the thermal capacitance are investigated.
In all of these calculations, a comparison is made with an effective
medium theory which becomes increasingly accurate as the structure
sizes fall well below the relevant resonance wavelength. Results show
that new levels of control over the magnitude and spectral contributions
to thermal capacitance can be achieved with corrugated structures
relative to planar ones.
\end{abstract}
\maketitle

\section*{Introduction:}

The control of thermal emission is critical to a variety of applications
such as energy conversion \citep{basu2007microscale,lin2003three},
imaging \citep{de2006thermal} and thermal emitters \citep{schuller2009optical,PhysRevB.76.045427}.
One way to achieve control over the thermal emission is obtained by
manipulating near-field surrounding optically-resonant nanostructures
\citep{shitrit2013spin,PhysRevLett.105.136402}. Radiative thermal
transfer between two objects which obeys Planck's law \citep{planck1914theory}
in the far field limit, show a dramatic enhancement when the separation
is reduced to such an extent that near-field effects dominate the
thermal transfer \citep{polder1971theory,shchegrov2000near}. Near
field effects cause a redistribution of the local density of states
(LDOS) and enable evanescent waves to make the most significant contribution
to the total thermal transfer. In addition to the total magnitude
of the thermal transfer, the spectral contributions also dramatically
change in the near field regime \citep{shchegrov2000near}. 

Recent developments in area of nanophotonics have inspired efforts
to use structures with subwavelength features for the purpose of controlling
radiative thermal transfer. An exact theory is available to quantify
the thermal transfer between arbitrary number and arbitrary shape
of materials \citep{kruger2012trace}. However, finding numerical
solutions to seemingly simple geometries (e.g. a nanoparticle above
a plane) require tremendous computational power as multiple frequencies
and length scales are involved. For this reason, there has been intense
efforts in this area to develop new, efficient numerical techniques
that enable calculation of thermal transfer in specific geometries.
This enabled calculation of thermal transfer in important basic geometries,
such as planar-to-planar \citep{polder1971theory} as well as planar
structures to a sphere \citep{PhysRevLett.106.210404,PhysRevB.84.245431,PhysRevB.85.165104},
a cylinder \citep{PhysRevB.85.165104}, and even a cone \citep{PhysRevB.85.165104}.
A good review that summarizes the results for these and other is given
in reference \citep{reid2013fluctuation}. 

In addition to the development of faster numerical techniques, physical
insight is also used to improve the speed by making certain reasonable
approximations. For example, effective medium theory has been used
to speed up calculation of the thermal transfer between subwavelength
periodic structures \citep{biehs2011modulation,basu2013near,biehs2012hyperbolic,biehs2013super,Biehs2011}.
This theory transforms high spatial frequency structures to uniform,
simple structures for which the variation in optical properties happen
just along a single dimension creating a stratified medium; After
that, theories to deal with stratified media \citep{polder1971theory}
can be applied for calculation of the thermal transfer. Effective
medium theories can not handle periodic structures with structure
sizes and spacings that are not deep subwavelength for all of the
relevant wavelengths in the problem. Here the relevant wavelengths
can be linked to materials-related resonances (e.g. plasmonic or phononic)
or structure related resonances (e.g. Mie or grating resonances).

In this paper we theoretically derive an expression for the radiative
thermal heat transfer in periodic structures based on rigorous coupled
wave analysis (RCWA) method that can handle such structures. This
enables one to access new physical regimes and to discover and systematically
analyze new physical phenomena in thermal transfer physics. Even tough
thermal emission from periodic structures to air is investigated in
several references \citep{Wang2012,han2009theory}, to our knowledge
this is the first time, such a theory is developed for rigorously
obtaining thermal transfer between nano-structured periodic materials
and a planar structure in the near field regime. The RCWA technique
together with the possible use of symmetries in the systems boosts
the numerical efficiency compared with the simulations that has been
done for calculation of thermal transfer between grating structures
using the FDTD method, recently \citep{PhysRevLett.107.114302}. 

The RCWA formalism, provides significant flexibility to include arbitrary-shaped
nanostructures and good criteria for determining the accuracy of obtained
results based on convergence by increasing the number of spatial harmonics.
Our method has some resemblance to the scattering method \citep{Bimonte2009,PhysRevLett.106.210404}
in its final form, however, there are some distinguishing technical
differences. Our method also provides a very direct way for determining
the variation of thermal transfer across a period in the periodic
structures. This variation can itself give important information to
determine whether the periodicity is in the subwavelength regime or
not. For instance, in the regime that periodicity is on the same order
or even larger than the resonance wavelength, we expect the thermal
transfer flow should be maximum in the regions that the top and bottom
layer are closer together and vice versa. In fact, in this regime,
total thermal transfer can be seen as a superposition of two parallel
channels, one with smaller value coming from the regions with larger
gap size and the other one with a larger value coming from the regions
with smaller gap size. This decomposition breaks down in the regime
that periodicity becomes subwavelength, in which effective medium
theory becomes more accurate, and the cross talk between two adjacent
regions become increasingly important. In the deep subwavelength regime,
thermal transfer should have negligible variation across the period.

Use of the RCWA method for obtaining electromagnetic field patterns
is quite common in nanophotonics. A numerically stable version of
this method was first developed by moharam \citep{moharam1995formulation,moharam1995stable},
and this technique can be used to obtain electromagnetic field distributions
developed around arbitrary periodic structures under plane wave incident
field illumination. However, for thermal transfer calculations we
will use it to calculate the green function that captures the electromagnetic
field response to arbitrary located and oriented electric dipoles.
For calculation of the Green function with the RCWA method we have
made use of the modified Sipe\textquoteright{}s formalism \citep{sipe1987new,han2009theory}. 

In continuation, the derived method is used for calculation of the
thermal capacitance between two SiC slabs, where one of them is patterned
with a grating structure at different values of the duty cycle. The
thermal capacitance is also calculated between a SiC slab and an array
of SiC beams of rectangular cross section. Here, the dependence of
the thermal transfer on beam size is explored. SiC is a polar semiconductor
and its surface supports resonant collective lattice vibrations known
as surface phonon polaritons (SPPs). These resonances which are in
the infrared region, provide the main channels for thermal transfer
in the near-field regime. The numerical calculations have been done
for spacings and periodicities that span several orders of magnitudes
to explore different physical regimes for the thermal transport. Since
SiC has a resonance wavelength around 10$\mu$m, we also expect Mie
resonances to show up themselves in these range of distances. Our
calculations verify this hypothesis by showing that in this range
of distances, the thermal capacitance obtains its extremum magnitudes
for non planar structures. This observation verifies that periodic
structures can be used to reach new levels of control over thermal
transfer and access new resonant pathways that enhance or spectrally
control the thermal transfer.

\section*{Theory:}

\begin{figure}
\begin{centering}
\subfloat[]{\centering{}\includegraphics[scale=0.7]{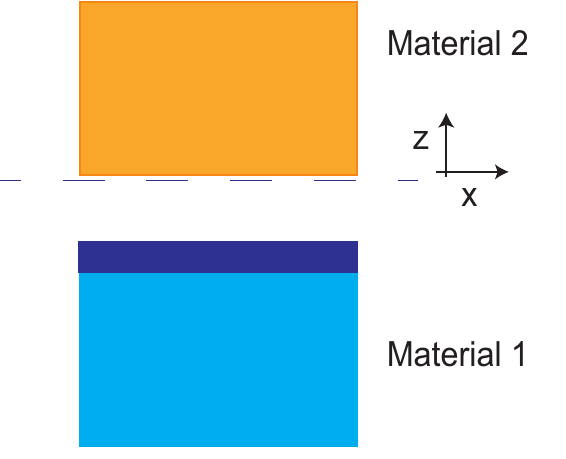}\label{planar_structure_fig}}\subfloat[]{\centering{}\includegraphics[scale=0.7]{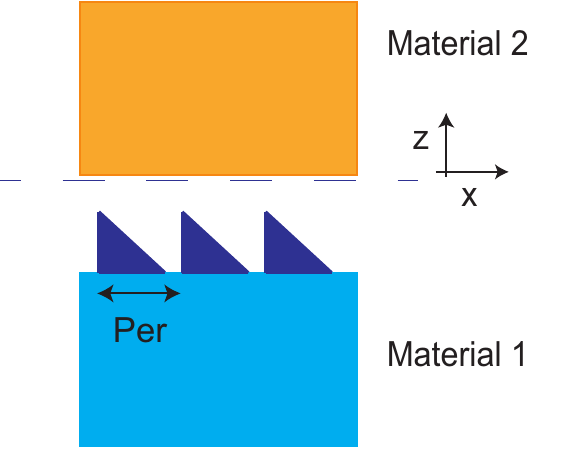}\label{periodic_structure_fig}}
\par\end{centering}

\centering{}\caption{Schematic of (a) planar structured materials and (b) a planar and
an arbitrary periodic shaped structure that will be analyzed in thermal
transfer calculations}
\label{schematics_fig}
\end{figure}

Before deriving the theory used for calculating the thermal transfer
from a periodic to a planar structure, it is educational to review
the derivation of Green functions in planar structures through the
use of Sipe\textquoteright{}s method \citep{sipe1987new}. Thermal
transfer calculations involving planar structures was first done by
Van hove and Polder in 1971 \citep{polder1971theory}. Sipe showed
how the required Green functions for calculation of thermal transfer
can easily be re-derived in a convenient form for an arbitrary stack
of planar materials. The first sub-section of this part is devoted
to this re-derivation. This corresponds to calculation of green function
in structures like the one shown in Fig. \ref{planar_structure_fig}. 

In the second part of this section, we apply the Sipe\textquoteright{}s
approach to obtain the Green function for periodic structures. We
will use this Green function later for obtaining the thermal transfer
through calculation of the Poynting vector that captures the thermal
power flow from one medium to another. A schematic of the type of
periodic structures of interest is illustrated in Fig. \ref{periodic_structure_fig}.
For illustration purposes and to simplify the math involved for this
case, we restrict ourselves to have one of the materials to be planar.

\subsection*{Green\textquoteright{}s function for stratified media:}

For the calculation of thermal transfer, we need to calculate the
Green function which determines the produced electromagnetic fields
in one object (say object 1) that result from current sources in the
other object (say object 2). For stratified media composed of a stack
of different layers, the contribution of an infinitesimal current
$\vec{J}\left(k_{x},k_{y}\right)dz^{\prime}$ to the total electric
field produced by it at its own position, denoted by $z^{\prime}$
in object 2, is given by: 

\begin{equation}
\overrightarrow{dE_{inc}}\left(z^{\prime}\right)=\frac{-\omega\mu_{0}}{2k_{z}}\left(\hat{p}_{2+}\hat{p}_{2+}+\hat{s}\hat{s}\right)\cdot\vec{J}dz^{\prime}
\end{equation}

where $\hat{p}_{2+}$ and $\hat{s}$ refers to the P and S polarization
directions for a wave with transverse wave-vector components of $k_{x},k_{y}$.
Note that $+$ sign denotes the wave with the wave-vector direction
from object 2 toward object 1. From the transfer matrix method, the
electric field produced by that element in a location different from
its own location, denoted by $z$ in material 1, is given by:

\begin{equation}
\overrightarrow{dE_{inc}}\left(z\right)=\frac{-\omega\mu_{0}}{2k_{z}}\left(t_{21}^{p}\hat{p}_{1+}\hat{p}_{2+}+t_{21}^{s}\hat{s}\hat{s}\right)\cdot\vec{J}dz^{\prime}
\end{equation}

where $\hat{p}_{1+}$ refers to P polarization direction in object
1, and $t_{21}^{s}$ and $t_{21}^{p}$ refers to transmission coefficients
from $z^{\prime}$ to $z$, for S and P polarizations, respectively. 

Accordingly, the dyadic Green function is given through the following
expression:

\begin{equation}
\overleftrightarrow{G}\left(z,z^{\prime}\right)=\frac{-\omega\mu_{0}}{2k_{z}}\left(t_{21}^{p}\hat{p}_{1+}\hat{p}_{2+}+t_{21}^{s}\hat{s}\hat{s}\right)
\end{equation}

This exactly follows Sipe\textquoteright{}s derivation for the Green
function. 

In the simple case of a slab adjacent to air, this formalism can be
simplified further. In this case, we assume the boundary between them
is located at $z=0$ and the observation point be located inside air
at $z=0$. Moreover, we assume that a current source to be located
inside the slab at a distance $z^{\prime}$ from the object's surface.

Denoting the transversal wave vector by $\beta\hat{\beta}=k_{x}\hat{x}+k_{y}\hat{y}$,
the total wave vector in the air and slab can be expressed as: 

\begin{align}
\vec{k_{1}} & =\beta\hat{\beta}+k_{zv}\hat{z}\\
\vec{k_{2}} & =\beta\hat{\beta}+k_{z}\hat{z}
\end{align}

Also the S and P polarization directions can be expressed as: 

\begin{align}
\hat{p}_{2+} & =\frac{c}{\sqrt{\epsilon}\omega}\left(\beta\hat{z}-k_{z}\hat{\beta}\right)\\
\hat{p}_{1+} & =\frac{c}{\omega}\left(\beta\hat{z}-k_{zv}\hat{\beta}\right)\\
\hat{s} & =\hat{\beta}\times\hat{z}
\end{align}

Moreover in this case, $t_{21}^{p}$ and $t_{21}^{s}$ are simple
Fresnel coefficients: 

\begin{align}
t_{21}^{p} & =\frac{2\sqrt{\epsilon}k_{z}}{k_{z}+\epsilon k_{zv}}e^{-ik_{z}z^{\prime}}\\
t_{21}^{s} & =\frac{2k_{z}}{k_{z}+k_{zv}}e^{-ik_{z}z^{\prime}}
\end{align}

So the dyadic Green function can easily be derived as:

\begin{align}
\overleftrightarrow{G}= & -\frac{\mu_{0}c^{2}e^{-ik_{z}z^{\prime}}}{\omega\left(k_{z}+\epsilon k_{zv}\right)}\left(\beta\hat{z}-k_{zv}\hat{\beta}\right)\left(\beta\hat{z}-k_{z}\hat{\beta}\right)\nonumber \\
 & -\frac{\mu_{0}\omega e^{-ik_{z}z^{\prime}}}{k_{z}+k_{zv}}\left(\hat{\beta}\times\hat{z}\right)\left(\hat{\beta}\times\hat{z}\right)
\end{align}

with the following components:

\begin{widetext}

\begin{equation}
\overleftrightarrow{G}=\begin{bmatrix}\frac{-k_{x}^{2}}{\omega\epsilon_{0}\beta^{2}}\frac{k_{zv}k_{z}e^{-ik_{z}z^{\prime}}}{k_{z}+\epsilon k_{zv}}-\frac{\omega\mu_{0}}{\beta^{2}}\frac{k_{y}^{2}e^{-ik_{z}z^{\prime}}}{k_{z}+k_{zv}} & \frac{-k_{x}k_{y}}{\omega\epsilon_{0}\beta^{2}}\frac{k_{zv}k_{z}e^{-ik_{z}z^{\prime}}}{k_{z}+\epsilon k_{zv}}+\frac{\omega\mu_{0}}{\beta^{2}}\frac{k_{x}k_{y}e^{-ik_{z}z^{\prime}}}{k_{z}+k_{zv}} & \frac{1}{\omega\epsilon_{0}}\frac{k_{zv}k_{x}e^{-ik_{z}z^{\prime}}}{k_{z}+\epsilon k_{zv}}\\
\frac{-k_{y}k_{x}}{\omega\epsilon_{0}\beta^{2}}\frac{k_{zv}k_{z}e^{-ik_{z}z^{\prime}}}{k_{z}+\epsilon k_{zv}}+\frac{\omega\mu_{0}}{\beta^{2}}\frac{k_{x}k_{y}e^{-ik_{z}z^{\prime}}}{k_{z}+k_{zv}} & \frac{-k_{y}^{2}}{\omega\epsilon_{0}\beta^{2}}\frac{k_{zv}k_{z}e^{-ik_{z}z^{\prime}}}{k_{z}+\epsilon k_{zv}}-\frac{\omega\mu_{0}}{\beta^{2}}\frac{k_{x}^{2}e^{-ik_{z}z^{\prime}}}{k_{z}+k_{zv}} & \frac{1}{\omega\epsilon_{0}}\frac{k_{y}k_{zv}e^{-ik_{z}z^{\prime}}}{k_{z}+\epsilon k_{zv}}\\
\frac{1}{\omega\epsilon_{0}}\frac{k_{x}k_{z}e^{-ik_{z}z^{\prime}}}{k_{z}+\epsilon k_{zv}} & \frac{1}{\omega\epsilon_{0}}\frac{k_{y}k_{z}e^{-ik_{z}z^{\prime}}}{k_{z}+\epsilon k_{zv}} & \frac{-\beta^{2}}{\omega\epsilon_{0}}\frac{e^{-ik_{z}z^{\prime}}}{k_{z}+\epsilon k_{zv}}
\end{bmatrix}
\end{equation}

\end{widetext}

These components become the more known results if $k_{y}$ is assumed
to be zero, as shown for instance in reference \citep{polder1971theory}:

\begin{align}
\overleftrightarrow{G} & =\nonumber \\
 & \begin{bmatrix}\frac{-1}{\omega\epsilon_{0}}\frac{k_{zv}k_{z}e^{-ik_{z}z^{\prime}}}{k_{z}+\epsilon k_{zv}} & 0 & \frac{1}{\omega\epsilon_{0}}\frac{k_{zv}k_{x}e^{-ik_{z}z^{\prime}}}{k_{z}+\epsilon k_{zv}}\\
0 & \frac{-\omega\mu_{0}e^{-ik_{z}z^{\prime}}}{k_{z}+k_{zv}} & 0\\
\frac{1}{\omega\epsilon_{0}}\frac{k_{x}k_{z}e^{-ik_{z}z^{\prime}}}{k_{z}+\epsilon k_{zv}} & 0 & \frac{-1}{\omega\epsilon_{0}}\frac{k_{x}^{2}e^{-ik_{z}z^{\prime}}}{k_{z}+\epsilon k_{zv}}
\end{bmatrix}
\end{align}

It is clear that the these results can be easily generalized to more
complicated planar structures by deriving more general expressions
for the Fresnel coefficients.

\subsection*{Generalization of the Green\textquoteright{}s function to periodic
structures:}

In the general case of periodic structures, we have:

\begin{align}
 & \overleftrightarrow{G_{E}}\left(\omega,x,y,z,\beta\hat{\beta},z^{\prime}\right)=\nonumber \\
 & -\frac{\omega\mu_{0}}{2k_{z}}\overrightarrow{Res_{E}}\left(\omega,x,y,z,\beta\hat{\beta},z^{\prime},\hat{p}_{2+}\right)\hat{p}_{2+}\nonumber \\
 & -\frac{\omega\mu_{0}}{2k_{z}}\overrightarrow{Res_{E}}\left(\omega,x,y,z,\beta\hat{\beta},z^{\prime},\hat{s}\right)\hat{s}
\end{align}

where $\overrightarrow{Res_{E}}\left(\omega,x,y,z,\beta\hat{\beta},z^{\prime},\hat{p}_{2+}\right)$
and $\overrightarrow{Res_{E}}\left(\omega,x,y,z,\beta\hat{\beta},z^{\prime},\hat{s}\right)$
are electric field responses at position $x,y,z$ to the P and S polarized
incident plane wave with transversal wave-vector $\beta\hat{\beta}$
and unity electric field amplitude at position $z^{\prime}$ and angular
frequency $\omega$. This is the modified version of sipe's formalism
\citep{sipe1987new}. 

Similarly, for the magnetic field, the following Green function is
defined:

\begin{align}
 & \overleftrightarrow{G_{H}}\left(\omega,x,y,z,\beta\hat{\beta},z^{\prime}\right)=\nonumber \\
 & -\frac{\omega\mu_{0}}{2k_{z}}\overrightarrow{Res_{H}}\left(\omega,x,y,z,\beta\hat{\beta},z^{\prime},\hat{p}_{2+}\right)\hat{p}_{2+}\nonumber \\
 & -\frac{\omega\mu_{0}}{2k_{z}}\overrightarrow{Res_{H}}\left(\omega,x,y,z,\beta\hat{\beta},z^{\prime},\hat{s}\right)\hat{s}
\end{align}

where $\overrightarrow{Res_{H}}\left(\omega,x,y,z,\beta\hat{\beta},z^{\prime},\hat{p}_{2+}\right)$
and $\overrightarrow{Res_{H}}\left(\omega,x,y,z,\beta\hat{\beta},z^{\prime},\hat{s}\right)$
are magnetic field responses at position $x,y,z$ to the P and S polarized
incident plane wave, again with transversal wave vector $\beta\hat{\beta}$
and unity electric field amplitude at position $z^{\prime}$ and angular
frequency of $\omega$. 

From these, in the general case of periodic structures, for the current
density of $\vec{J}\left(\omega,k_{x}^{\prime},k_{y}^{\prime},z^{\prime}\right)=\vec{J}\left(\omega,k_{x},k_{y},z_{0}\right)\delta\left(k_{x}^{\prime}-k_{x}\right)\delta\left(k_{y}^{\prime}-k_{y}\right)\delta\left(z^{\prime}-z_{0}\right)$,
the generated electric and magnetic field components at position $x,y$
and $z=0$ are given by:

\begin{align}
 & \overrightarrow{E_{a}}\left(\omega,x,y,z=0,k_{x},k_{y},z_{0}\right)=\frac{-\omega\mu_{0}}{2k_{z}}e^{-ik_{z}\left(k_{x},k_{y}\right)z_{0}}\nonumber \\
 & \times\sum_{b}\Big(\overrightarrow{Res_{E}}\left(\omega,x,y,\beta\hat{\beta},\hat{p}_{2+}\right)\hat{p}_{2+}\nonumber \\
 & +\overrightarrow{Res_{E}}\left(\omega,x,y,\beta\hat{\beta},\hat{s}\right)\hat{s}\Big)_{ab}\vec{J}_{b}\left(\omega,k_{x},k_{y},z_{0}\right)
\end{align}

\begin{align}
 & \overrightarrow{H_{a}}\left(\omega,x,y,z=0,k_{x},k_{y},z_{0}\right)=\frac{-\omega\mu_{0}}{2k_{z}}e^{-ik_{z}\left(k_{x},k_{y}\right)z_{0}}\nonumber \\
 & \times\sum_{b}\Big(\overrightarrow{Res_{H}}\left(\omega,x,y,\beta\hat{\beta},\hat{p}_{2+}\right)\hat{p}_{2+}\nonumber \\
 & +\overrightarrow{Res_{H}}\left(\omega,x,y,\beta\hat{\beta},\hat{s}\right)\hat{s}\Big)_{ab}\vec{J}_{b}\left(\omega,k_{x},k_{y},z_{0}\right)
\end{align}

where 

\begin{align}
 & \overrightarrow{Res_{E}}\left(\omega,x,y,\beta\hat{\beta},\hat{p}_{2+}\right)\triangleq\nonumber \\
 & \overrightarrow{Res_{E}}\left(\omega,x,y,z=0,\beta\hat{\beta},z^{\prime}=0,\hat{p}_{2+}\right)
\end{align}

\begin{align}
 & \overrightarrow{Res_{H}}\left(\omega,x,y,\beta\hat{\beta},\hat{p}_{2+}\right)\triangleq\nonumber \\
 & \overrightarrow{Res_{H}}\left(\omega,x,y,z=0,\beta\hat{\beta},z^{\prime}=0,\hat{p}_{2+}\right)
\end{align}

\begin{align}
 & \overrightarrow{Res_{E}}\left(\omega,x,y,\beta\hat{\beta},\hat{s}\right)\triangleq\nonumber \\
 & \overrightarrow{Res_{E}}\left(\omega,x,y,z=0,\beta\hat{\beta},z^{\prime}=0,\hat{s}\right)
\end{align}

\begin{align}
 & \overrightarrow{Res_{H}}\left(\omega,x,y,\beta\hat{\beta},\hat{s}\right)\triangleq\nonumber \\
 & \overrightarrow{Res_{H}}\left(\omega,x,y,z=0,\beta\hat{\beta},z^{\prime}=0,\hat{s}\right)
\end{align}

We assumed the z direction to be normal to plane of top slab. The
convention used for the x direction is also shown in Fig. \ref{schematics_fig}.
In the above equations, $z=0$ is chosen as the the place of the top
slab. In fact we are interested only in the calculation of electromagnetic
fields in this location; since by knowing the transverse components
of $\vec{E}$ and $\vec{H}$ field in this plane, we can calculate
the Poynting vector which determines the heat transfer. This plane
is shown with dashed line in Figs. \ref{planar_structure_fig} and
\ref{periodic_structure_fig}.

For simplification of the later equations, we define $\overrightarrow{G_{E}^{a}}\left(\omega,x,y,k_{x},k_{y}\right)$
and $\overrightarrow{G_{H}^{a}}\left(\omega,x,y,k_{x},k_{y}\right)$
as the following: 

\begin{align}
 & \overrightarrow{G_{E}^{a}}\left(\omega,x,y,k_{x},k_{y}\right)\triangleq\frac{-\omega\mu_{0}}{2k_{z}}\nonumber \\
 & \times\sum_{b}\Big(\overrightarrow{Res_{E}}\left(\omega,x,y,\beta\hat{\beta},\hat{p}_{2+}\right)\hat{p}_{2+}\nonumber \\
 & +\overrightarrow{Res_{E}}\left(\omega,x,y,\beta\hat{\beta},\hat{s}\right)\hat{s}\Big)_{ba}\hat{e}_{b}
\end{align}

\begin{align}
 & \overrightarrow{G_{H}^{a}}\left(\omega,x,y,k_{x},k_{y}\right)\triangleq\frac{-\omega\mu_{0}}{2k_{z}}\nonumber \\
 & \times\sum_{b}\Big(\overrightarrow{Res_{H}}\left(\omega,x,y,\beta\hat{\beta},\hat{p}_{2+}\right)\hat{p}_{2+}\nonumber \\
 & +\overrightarrow{Res_{H}}\left(\omega,x,y,\beta\hat{\beta},\hat{s}\right)\hat{s}\Big)_{ba}\hat{e}_{b}
\end{align}

where $\hat{e}_{b}$ is the unity vector in direction b, which takes
on the unity vectors in x,y, and z directions in the summation. These
are the electric and magnetic fields at position $x$, $y$, and $z=0$,
produced by the unity component $a$ of the current density at $z^{\prime}=0$.
Note that $\overrightarrow{Res_{E}}\left(\omega,x,y,\beta\hat{\beta},\hat{p}_{2+}\right)$,
$\overrightarrow{Res_{E}}\left(\omega,x,y,\beta\hat{\beta},\hat{s}\right)$,
$\overrightarrow{Res_{H}}\left(\omega,x,y,\beta\hat{\beta},\hat{p}_{2+}\right)$,
and $\overrightarrow{Res_{H}}\left(\omega,x,y,\beta\hat{\beta},\hat{s}\right)$
are the electromagnetic responses of the system that can be obtained
through the RCWA method. Consequently, $\overrightarrow{G_{E}^{a}}\left(\omega,x,y,k_{x},k_{y}\right)$
and $\overrightarrow{G_{H}^{a}}\left(\omega,x,y,k_{x},k_{y}\right)$
can be calculated directly from the RCWA method, as well.

Therefore, for a general current density distribution $\vec{J}\left(\omega,x_{0},y_{0},z_{0}\right)$
in the top material, we can write $\vec{E}$ and $\vec{H}$ at position
$x$, $y=0$, and $z=0$, in the following general form:

\begin{align}
 & \overrightarrow{E}\left(x,y=0,z=0,t\right)=\frac{1}{\left(2\pi\right)^{3}}\int_{\omega=0}^{+\infty}d\omega e^{i\omega t}\int\vec{dr_{0}}\nonumber \\
 & \times\sum_{a}\int_{k_{y}=-\infty}^{+\infty}\int_{k_{x}=-\infty}^{+\infty}dk_{x}dk_{y}e^{-ik_{x}x_{0}-ik_{y}y_{0}-ik_{z}\left(k_{x},k_{y}\right)z_{0}}\nonumber \\
 & \times\overrightarrow{G_{E}^{a}}\left(\omega,x,y=0,k_{x},k_{y}\right)\vec{J}_{a}\left(\omega,x_{0},y_{0},z_{0}\right)+\text{{c.c.}}
\end{align}

\begin{align}
 & \overrightarrow{H}\left(x,y=0,z=0,t\right)=\frac{1}{\left(2\pi\right)^{3}}\int_{\omega=0}^{+\infty}d\omega e^{i\omega t}\int\vec{dr_{0}}\nonumber \\
 & \times\sum_{b}\int_{k_{y}=-\infty}^{+\infty}\int_{k_{x}=-\infty}^{+\infty}dk_{x}dk_{y}e^{-ik_{x}x_{0}-ik_{y}y_{0}-ik_{z}\left(k_{x},k_{y}\right)z_{0}}\nonumber \\
 & \times\overrightarrow{G_{H}^{b}}\left(\omega,x,y=0,k_{x},k_{y}\right)\vec{J}_{b}\left(\omega,x_{0},y_{0},z_{0}\right)+\text{{c.c.}}
\end{align}

where $a$, $b$ denotes the three possible components of the current
density, and $\overrightarrow{G_{E}^{a}}\left(\omega,x,y=0,k_{x},k_{y}\right)$
and $\overrightarrow{G_{H}^{b}}\left(\omega,x,y=0,k_{x},k_{y}\right)$,
are defined in the above. Also $\text{{c.c.}}$ refers to complex
conjugate.

According to the above equations, the following expression for the
Poynting vector is found:

\begin{align}
 & \overrightarrow{P}\left(x,y=0,z=0\right)=\frac{1}{\left(2\pi\right)^{6}}\sum_{a,b}\int_{\omega=0}^{+\infty}\int_{\omega^{\prime}=0}^{+\infty}d\omega^{\prime}d\omega\nonumber \\
 & \times e^{i\left(\omega-\omega^{\prime}\right)t}\int\int\vec{dr_{0}}\vec{dr_{0}^{\prime}}\left\langle \vec{J}_{a}\left(\omega,\vec{r_{0}}\right)\vec{J}_{b}^{*}\left(\omega^{\prime},\vec{r_{0}^{\prime}}\right)\right\rangle \nonumber \\
 & \times\int_{k_{y}^{\prime}=-\infty}^{+\infty}\int_{k_{x}^{\prime}=-\infty}^{+\infty}\int_{k_{y}=-\infty}^{+\infty}\int_{k_{x}=-\infty}^{+\infty}dk_{x}dk_{y}dk_{x}^{\prime}dk_{y}^{\prime}\nonumber \\
 & \times\left(\overrightarrow{G_{E}^{a}}\left(\omega,x,y=0,k_{x},k_{y}\right)\times\overrightarrow{G_{H}^{b*}}\left(\omega^{\prime},x,y=0,k_{x}^{\prime},k_{y}^{\prime}\right)\right)\nonumber \\
 & \times e^{-ik_{x}x_{0}-ik_{y}y_{0}-ik_{z}z_{0}+ik_{x}^{\prime}x_{0}^{\prime}+ik_{y}^{\prime}y_{0}^{\prime}+ik_{z}^{\prime*}z_{0}^{\prime}}+\text{{c.c.}}
\end{align}

Random thermal motions of charges inside a material, generate fluctuating
current densities. These current densities, for a material that is
in the thermodynamic equilibrium at temperature T, obey the following
correlation relation known as fluctuation dissipation theorem \citep{Eckhardt1982305,Joulain200559}:

\begin{align}
 & \left\langle \vec{J}_{a}\left(\omega,\vec{r_{0}}\right)\vec{J}_{b}^{*}\left(\omega^{\prime},\vec{r_{0}^{\prime}}\right)\right\rangle =4\pi\epsilon_{0}\epsilon^{\prime\prime}\left(\omega\right)\hbar\omega^{2}\nonumber \\
 & \times\left(e^{\hbar\omega/k_{b}T}-1\right)^{-1}\delta_{ab}\delta\left(\omega-\omega^{\prime}\right)\delta\left(\vec{r_{0}}-\vec{r_{0}^{\prime}}\right)
\end{align}

After making a simplification using the fluctuation dissipation theorem,
we have:

\begin{align}
 & \overrightarrow{P}\left(x,y=0,z=0\right)=\frac{1}{16\pi^{5}}\sum_{a}\int_{\omega=0}^{+\infty}d\omega\int\vec{dr_{0}}\epsilon_{0}\nonumber \\
 & \times\epsilon^{\prime\prime}\left(\omega\right)\hbar\omega^{2}\left(e^{\hbar\omega/k_{b}T}-1\right)^{-1}\nonumber \\
 & \times\int_{k_{y}^{\prime}=-\infty}^{+\infty}\int_{k_{x}^{\prime}=-\infty}^{+\infty}\int_{k_{y}=-\infty}^{+\infty}\int_{k_{x}=-\infty}^{+\infty}dk_{x}dk_{y}dk_{x}^{\prime}dk_{y}^{\prime}\nonumber \\
 & \times\left(\overrightarrow{G_{E}^{a}}\left(\omega,x,y=0,k_{x},k_{y}\right)\times\overrightarrow{G_{H}^{a*}}\left(\omega,x,y=0,k_{x}^{\prime},k_{y}^{\prime}\right)\right)\nonumber \\
 & \times e^{i\left(k_{x}^{\prime}-k_{x}\right)x_{0}+i\left(k_{y}^{\prime}-k_{y}\right)y_{0}}e^{i\left(k_{z}^{\prime*}-k_{z}\right)z_{0}}+\text{{c.c.}}
\end{align}

after interchanging the order of integrations, we arrive at:

\begin{align}
 & \overrightarrow{P}\left(x,y=0,z=0\right)=\frac{1}{16\pi^{5}}\sum_{a}\int_{\omega=0}^{+\infty}d\omega\epsilon_{0}\epsilon^{\prime\prime}\left(\omega\right)\nonumber \\
 & \times\int_{k_{y}^{\prime}=-\infty}^{+\infty}\int_{k_{x}^{\prime}=-\infty}^{+\infty}\int_{k_{y}=-\infty}^{+\infty}\int_{k_{x}=-\infty}^{+\infty}dk_{x}dk_{y}dk_{x}^{\prime}dk_{y}^{\prime}\nonumber \\
 & \times\left(\overrightarrow{G_{E}^{a}}\left(\omega,x,y=0,k_{x},k_{y}\right)\times\overrightarrow{G_{H}^{a*}}\left(\omega,x,y=0,k_{x}^{\prime},k_{y}^{\prime}\right)\right)\nonumber \\
 & \times\hbar\omega^{2}\int_{z_{0}=0}^{\infty}\int_{y_{0}=-\infty}^{\infty}\int_{x_{0}=-\infty}^{\infty}dx_{0}dy_{0}dz_{0}e^{i\left(k_{z}^{\prime*}-k_{z}\right)z_{0}}\nonumber \\
 & \times e^{i\left(k_{x}^{\prime}-k_{x}\right)x_{0}+i\left(k_{y}^{\prime}-k_{y}\right)y_{0}}\left(e^{\hbar\omega/k_{b}T}-1\right)^{-1}+\text{{c.c.}}
\end{align}

which reduces to:

\begin{align}
 & \overrightarrow{P}\left(x,y=0,z=0\right)=\frac{1}{4\pi^{3}}\sum_{a}\int_{\omega=0}^{+\infty}d\omega\epsilon_{0}\epsilon^{\prime\prime}\left(\omega\right)\nonumber \\
 & \times\int_{k_{y}^{\prime}=-\infty}^{+\infty}\int_{k_{x}^{\prime}=-\infty}^{+\infty}\int_{k_{y}=-\infty}^{+\infty}\int_{k_{x}=-\infty}^{+\infty}dk_{x}dk_{y}dk_{x}^{\prime}dk_{y}^{\prime}\nonumber \\
 & \times\left(\overrightarrow{G_{E}^{a}}\left(\omega,x,y=0,k_{x},k_{y}\right)\times\overrightarrow{G_{H}^{a*}}\left(\omega,x,y=0,k_{x},k_{y}\right)\right)\nonumber \\
 & \times\hbar\omega^{2}\left(e^{\hbar\omega/k_{b}T}-1\right)^{-1}\delta\left(k_{x}-k_{x}^{\prime}\right)\delta\left(k_{y}-k_{y}^{\prime}\right)\nonumber \\
 & \times\int_{z_{0}=0}^{\infty}e^{i\left(k_{z}^{\prime*}-k_{z}\right)z_{0}}dz_{0}+\text{{c.c.}}
\end{align}

Finally we obtain that:

\begin{align}
 & \overrightarrow{P}\left(z=0,y=0,x\right)=\frac{1}{4\pi^{3}}\sum_{a}\int_{\omega=0}^{+\infty}d\omega\epsilon_{0}\epsilon^{\prime\prime}\left(\omega\right)\hbar\omega^{2}\nonumber \\
 & \times\left(e^{\hbar\omega/k_{b}T}-1\right)^{-1}\int_{k_{y}=-\infty}^{+\infty}\int_{k_{x}=-\infty}^{+\infty}dk_{x}dk_{y}\nonumber \\
 & \times\frac{1}{Im\left(k_{z}\right)}\Re\Big\{\Big(\overrightarrow{G_{E}^{a}}\left(\omega,x,y=0,k_{x},k_{y}\right)\nonumber \\
 & \times\overrightarrow{G_{H}^{a*}}\left(\omega,x,y=0,k_{x},k_{y}\right)\Big)\Big\}
\end{align}

Note that the z component of this quantity measures the thermal transfer.
The thermal capacitance can be obtained from it through differentiating
with respect to temperature: 

\begin{align}
 & S_{total}(x)=\frac{1}{4\pi^{3}}\sum_{a}\int_{\omega=0}^{+\infty}d\omega\epsilon_{0}\epsilon^{\prime\prime}\left(\omega\right)\left(e^{\hbar\omega/k_{b}T}-1\right)^{-2}\nonumber \\
 & \times\int_{k_{y}=-\infty}^{+\infty}\int_{k_{x}=-\infty}^{+\infty}dk_{x}dk_{y}\frac{e^{\hbar\omega/k_{b}T}}{Im\left(k_{z}\right)}\frac{\hbar^{2}\omega^{3}}{k_{b}T^{2}}\nonumber \\
 & \times\Re\Big\{\Big(\overrightarrow{G_{E}^{a}}\left(\omega,x,y=0,k_{x},k_{y}\right)\nonumber \\
 & \times\overrightarrow{G_{H}^{a*}}\left(\omega,x,y=0,k_{x},k_{y}\right)\Big)_{z}\Big\}
\end{align}

In fact, what is measured as the total heat transfer and the corresponding
heat capacitance is the average of the above functions across a period,
which we show here with the same symbol:

\begin{equation}
S_{total}=\frac{1}{Per}\int_{x=0}^{Per}S_{total}\left(x\right)dx
\end{equation}

It is important to note that considering only the fields in a line
in x direction saves significant computational time. In fact this
is achieved by exploiting the translational symmetry of our structure
in y-direction and also the fact that for obtaining the energy flow,
it is sufficient to calculate the Poynting vector in a cross section.
Note that in our method, the variation of thermal transfer across
a cross section can also be obtained. This provides the ability to
determine the contributions of different locations across the period
to the total thermal transfer or capacitance. 

Moreover, in the above calculations, we are involved with only transverse
components of the electromagnetic fields. Since these quantities are
continuous across the barrier, we need only to calculate the electromagnetic
fields in the top material right at the boundary. These fields can
be calculated from the reflection coefficients in the RCWA formalism,
and will further simply the required RCWA calculations. In fact calculation
of the electromagnetic fields in the middle layers and bottom material
(transmission coefficients) are not needed anymore. {[}See section
7 of reference \citep{moharam1995stable}{]}

\section*{Numerical Results:}

\begin{figure}
\begin{centering}
\includegraphics[scale=0.8]{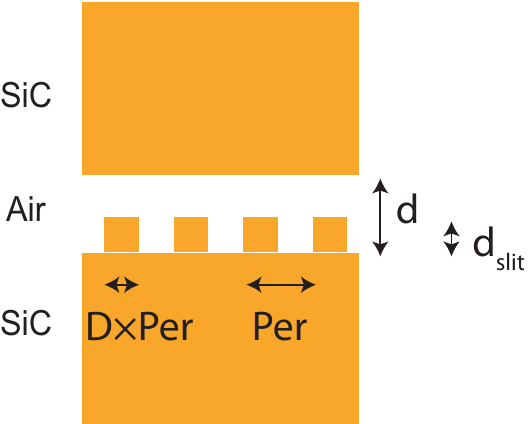}
\par\end{centering}

\centering{}\caption{Two SiC slab are placed in front of each other, one with a flat and
one with a binary grating structure. The duty cycle of the grating
is assumed to be $D$, and its periodicity is shown with $Per$. The
distances involved for this structure are shown in the figure. }
\label{grating_fig}
\end{figure}

\begin{figure}
\begin{centering}
\includegraphics[scale=0.8]{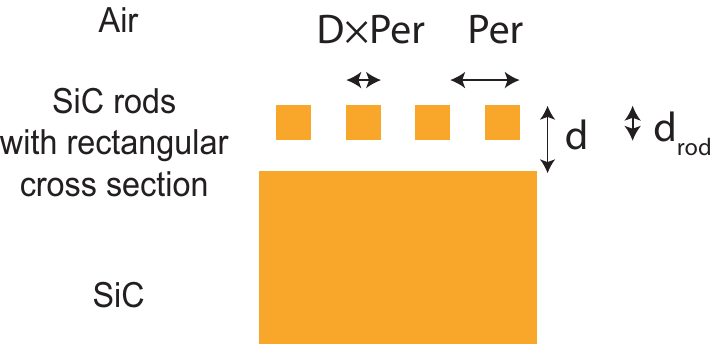}
\par\end{centering}

\centering{}\caption{SiC beams with a rectangular cross section are placed in front of
a SiC slab. The width of each beam is assumed to be $D\times Per$,
and they are separated by a distance $Per$ from each other. The distances
involved for this structure are shown in the figure.}
\label{rods_fig}
\end{figure}

In the following, we analyze two different periodic structures with
our developed formalism. The first example structure is shown in Fig.
\ref{grating_fig}. In this example, we modified the previously-studied
case of two closely-spaced SiC slabs separated from each other by
an air gap to a case where one of slabs is patterned with a periodic
grating. We expect that the corrugations modify the dispersion relation
of the surface phonon polaritons supported by a smooth SiC surface
and thereby impact the thermal capacitance . The second structure
that we considered, is an array of SiC beams of rectangular cross
section placed above a continuous slab of SiC (Fig. \ref{rods_fig}). 

Calculations for the first structure have been done for three different
separations between the two SiC structures (specifically $d=1um$,
$d=10um$, and $d=0.1um$) and different periodicities (specifically
$Per=1um$, $Per=10um$, and $Per=0.1um$). The depth of the grooves
in the considered structures is $d_{slit}=0.5d$.

The optical properties of the SiC material is computed based on references
\citep{spitzer1959infrared,spitzer1959infrared2} which assumes the
following expression for the SiC refractive index and extinction coefficient:

\begin{align}
n^{2} & =\frac{1}{2}\left\{ \left[\epsilon^{2}+4\left(\frac{\sigma}{\nu}\right)^{2}\right]^{\frac{1}{2}}+\epsilon\right\} \\
k^{2} & =\frac{1}{2}\left\{ \left[\epsilon^{2}+4\left(\frac{\sigma}{\nu}\right)^{2}\right]^{\frac{1}{2}}-\epsilon\right\} 
\end{align}

where the variables introduced in it are defined as:

\begin{align}
\epsilon & =\epsilon_{0}+4\pi\chi\\
\chi & =\rho\frac{1-\nu^{2}/\nu_{0}^{2}}{\left(1-\nu^{2}/\nu_{0}^{2}\right)^{2}+\gamma^{2}\nu^{2}/\nu_{0}^{2}}\\
\frac{\sigma}{\nu} & =2\pi\rho\frac{\gamma\nu}{\left(1-\nu^{2}/\nu_{0}^{2}\right)^{2}+\gamma^{2}\nu^{2}/\nu_{0}^{2}}
\end{align}

Containing the following numerical parameters:

\begin{align}
 & \rho=0.263,\nonumber \\
 & \gamma=0.006,\nonumber \\
 & \nu=2.38\times10^{13}sec^{-1}(12.6\mu m),\, and\nonumber \\
 & \epsilon_{0}=6.7
\end{align}

In addition, the temperature that is assumed in the numerical calculations
is $T=315K$.

For calculations based on the RCWA method, it is well-known that increasing
the number of harmonics leads to a more accurate determination of
the field distributions. However, this increase will lead to an increase
in computational time as well. In fact since the numerical evaluation
of the thermal capacitance by the presented method involves inverting
$4n\times4n$ matrices, the computational time grows with the cube
of the number of harmonics incorporated. It is clear from the last
equation in theory section that obtaining the spectral thermal capacitance
at a specific frequency requires two dimensional integrations in the
$k_{x}$,$k_{y}$ plane. For each value of $k_{x}$,$k_{y}$, a RCWA
calculation should be carried out to obtain the corresponding integrand.
This clarifies the importance of identifying a fast integration technique
to maximize the speed of calculations. We have used the VEGAS method
for integration in $k_{x}$,$k_{y}$ plane which is based on Monte
Carlo important sampling of the integrand function \citep{PeterLepage1978}.
To verify our calculation technique, we first accurately reproduced
the results for the limiting cases of gratings with duty cycles of
0 and 1. In those cases, using just one harmonic will lead to the
precise result and the RCWA method will converge to the results that
can be obtained with the transfer matrix method for a stratified medium
consisting of uniform layers. In these extremum cases we can simply
use the planar methods developed by Polder and Van hove \citep{polder1971theory}. 

\begin{figure}
\begin{centering}
\subfloat[]{\centering{}\includegraphics[scale=0.3]{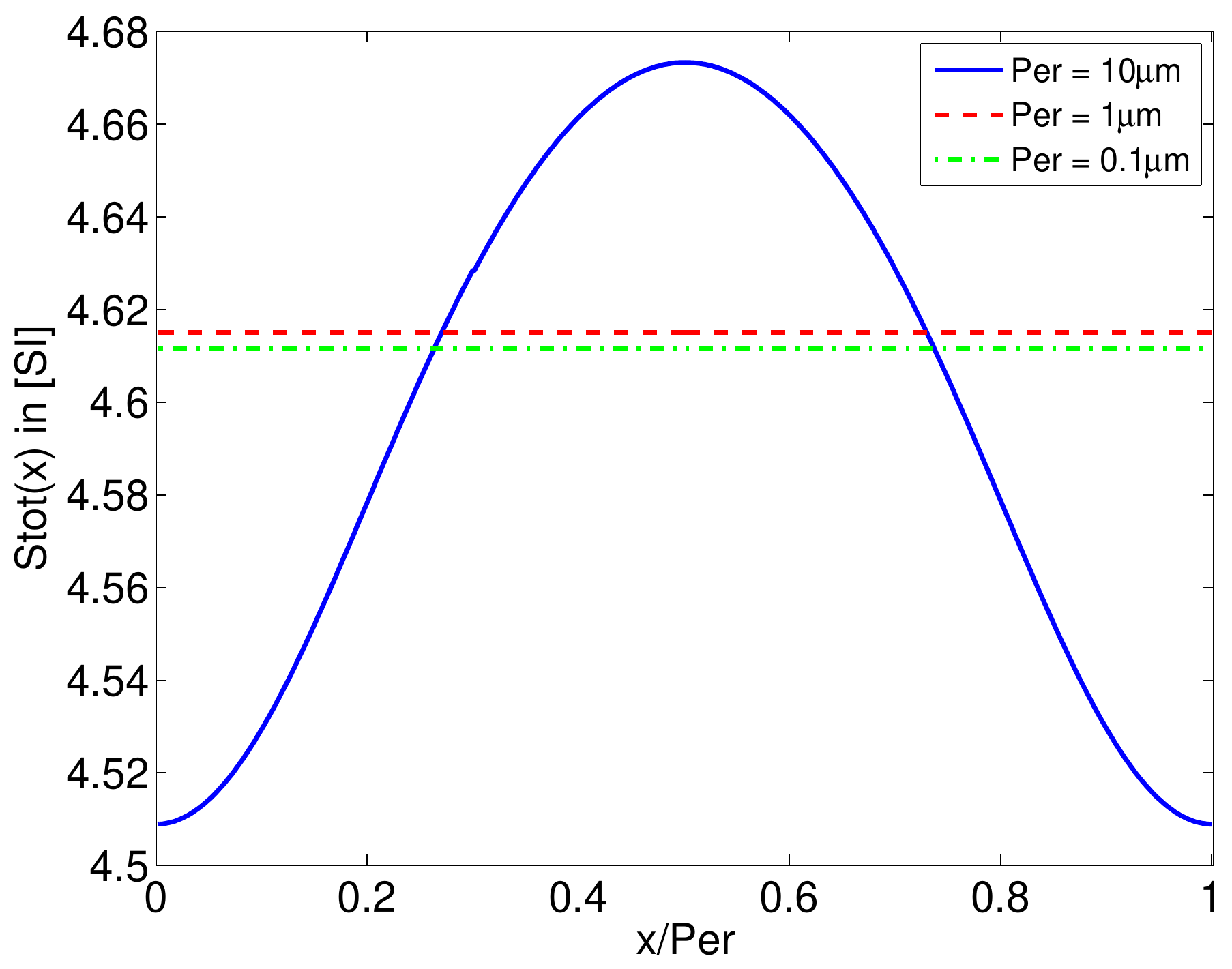}}\\
\subfloat[]{\centering{}\includegraphics[scale=0.3]{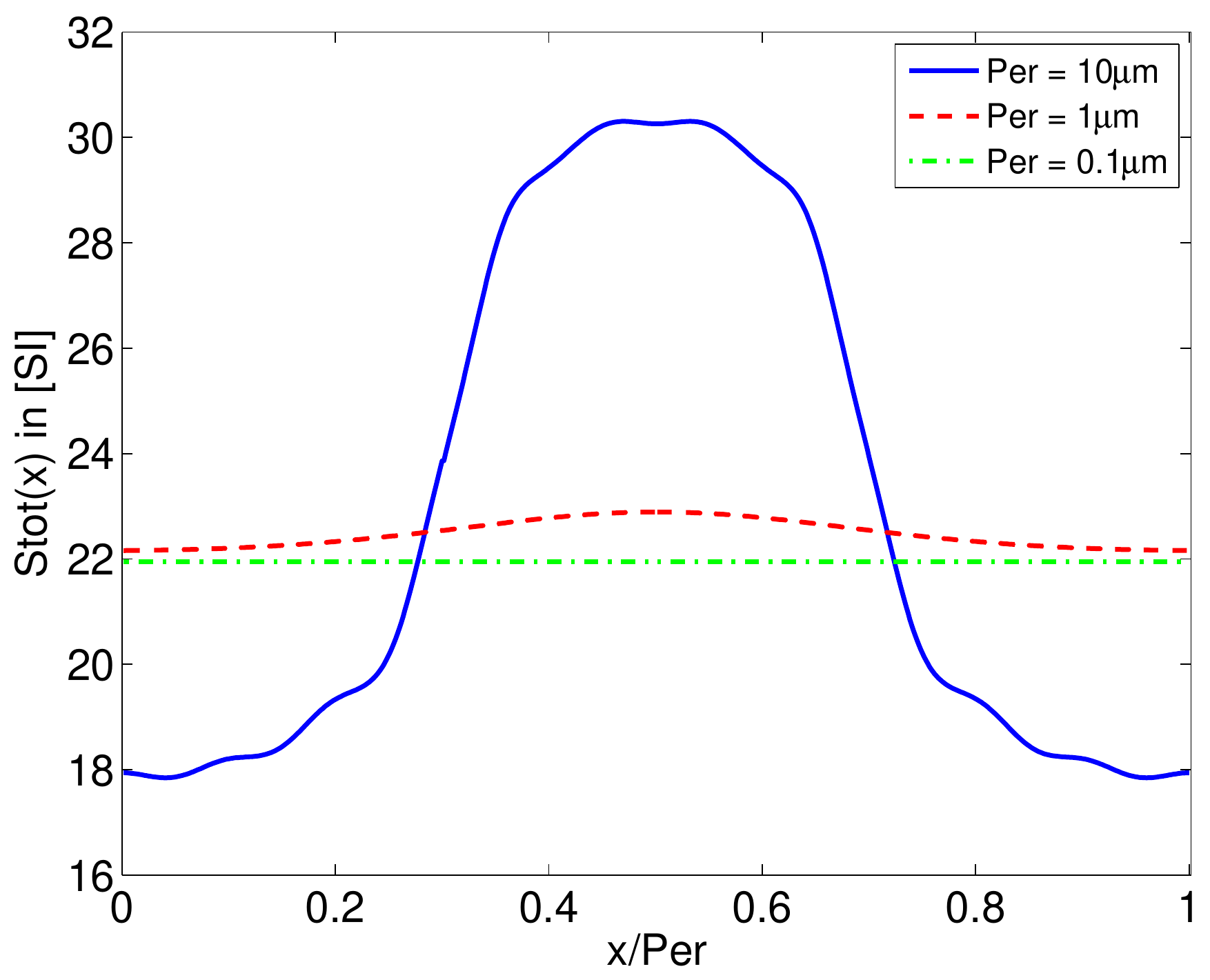}}\\
\subfloat[]{\centering{}\includegraphics[scale=0.3]{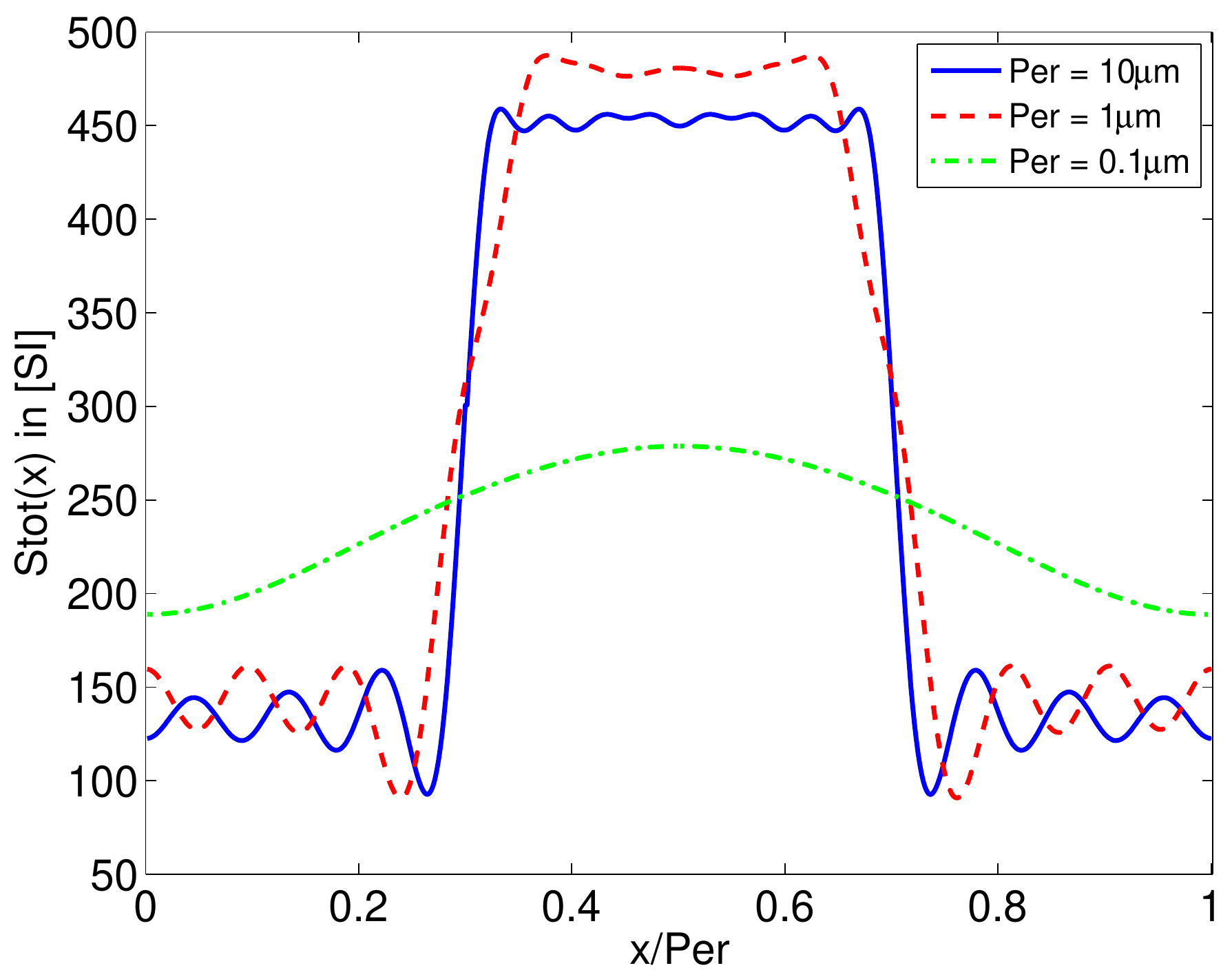}}
\par\end{centering}

\caption{Contributions to the thermal capacitance for the structure shown in
Fig. \ref{grating_fig} across a period for different values of periodicity,
and assuming a constant duty cycle of $D=0.4$, in the case of (a)
$d=10\mu m$ (b) $d=1\mu m$ (c) $d=0.1\mu m$}
\label{d_xvar_fig}
\end{figure}

To study the convergence of the results with the number of harmonics,
calculations were made with 4 different numbers of harmonics: 1, 5,
11, and 21. Obtained results show that for the considered structures,
the thermal capacitance converges with less than 2\% error without
the need for incorporating more harmonics. One important note regarding
our method is that this method in the case of incorporating just one
harmonic, $n=1$, reproduces the results obtained by using the effective
medium theory. Note that in the case of using just one harmonic, permittivity
of each layer is replaced by a constant value across the period. This
constant value, however, takes on different magnitudes depending on
the incident electric field direction. This is the case also in the
effective medium theories \citep{biehs2011modulation,basu2013near,biehs2012hyperbolic,biehs2013super,Biehs2011},
used for calculation of the thermal transfer, in which effective permittivities
of different layers are calculated as constant tensorial quantities.
In this regard, our method can be used to determine the accuracy of
the effective medium theory and how the actual responses are deviating
from it.

For this study, these numerical calculations were run on a node with
16 CPU s, using MPI \citep{gropp1999using} for parallelization (The
node that we used for our calculations has 16 processors of 2.67 GHz
Intel Xeon X5550). The time required for obtaining each set of results
on a single node for the case of 21 harmonics was around 10 hours.
However, this can be decreased by capitalizing on certain symmetries
in specific periodic structures, which has been proposed for the 2D
grating in reference \citep{bai2005group} and can be incorporated
in 1D grating structures as well (using for instance the inversion
symmetry present in the binary grating). 

\begin{figure*}
\begin{centering}
\subfloat[]{\includegraphics[scale=0.35]{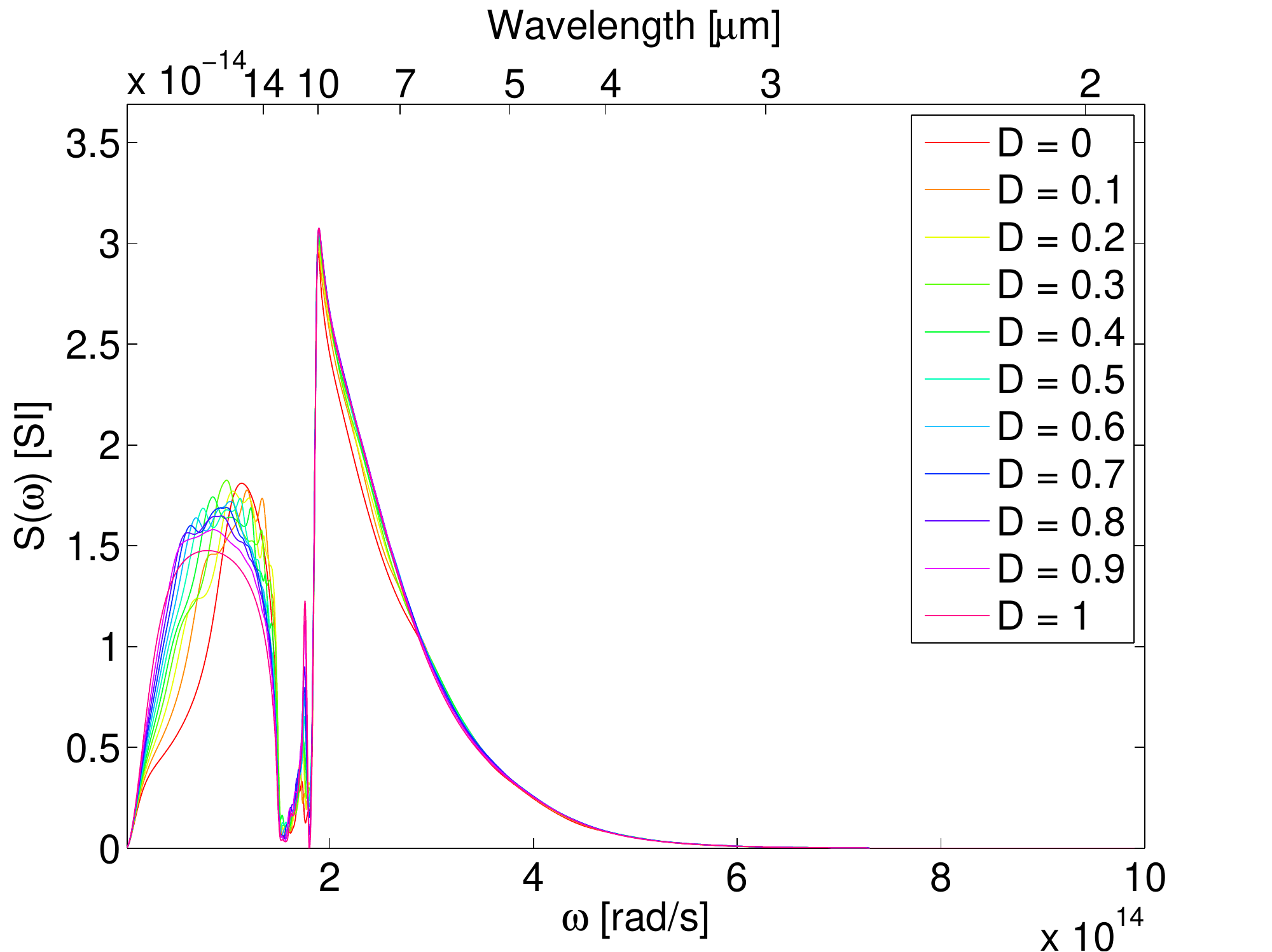}\includegraphics[scale=0.35]{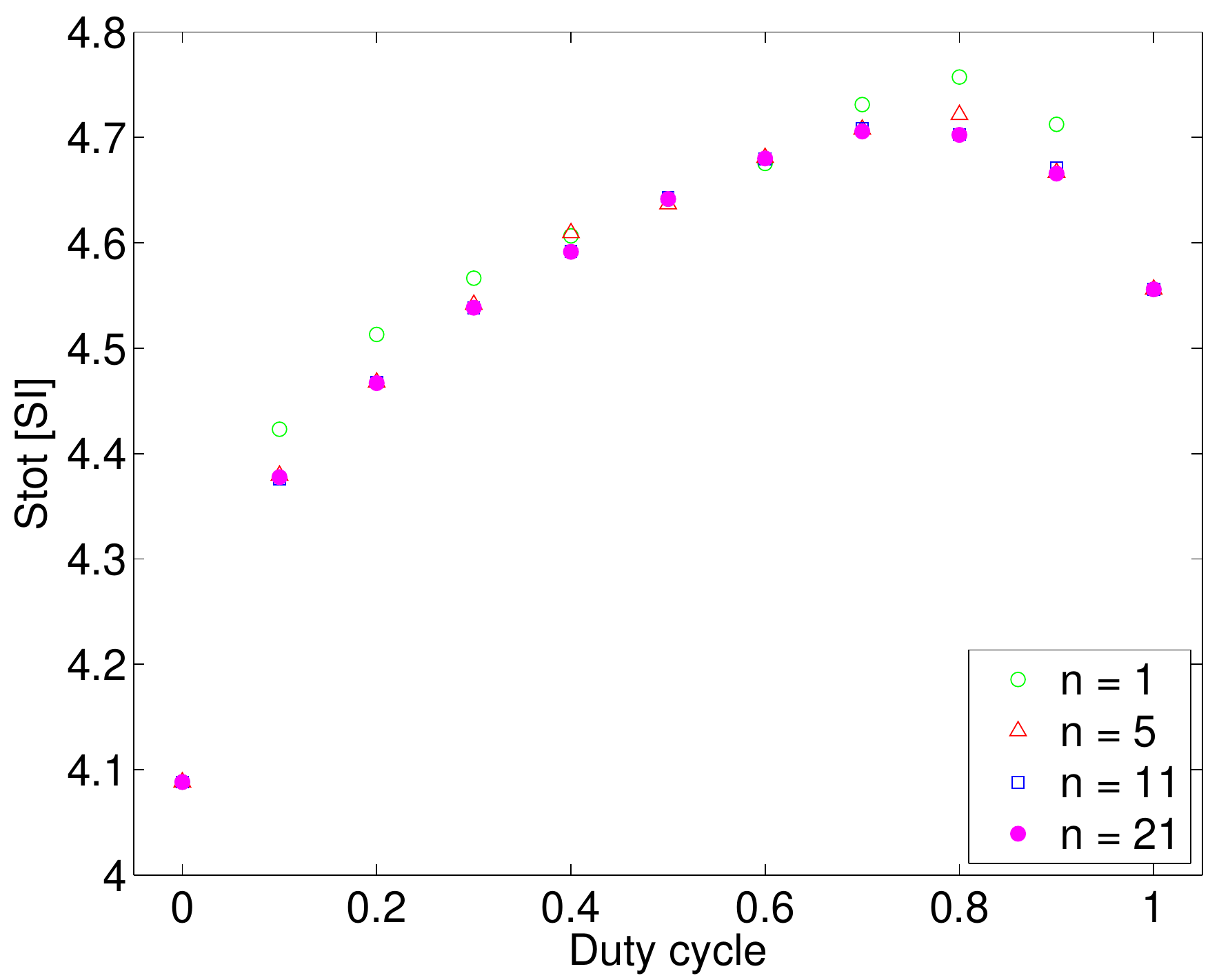}\label{d_10um_Per_10um_fig}}\\
\subfloat[]{\includegraphics[scale=0.35]{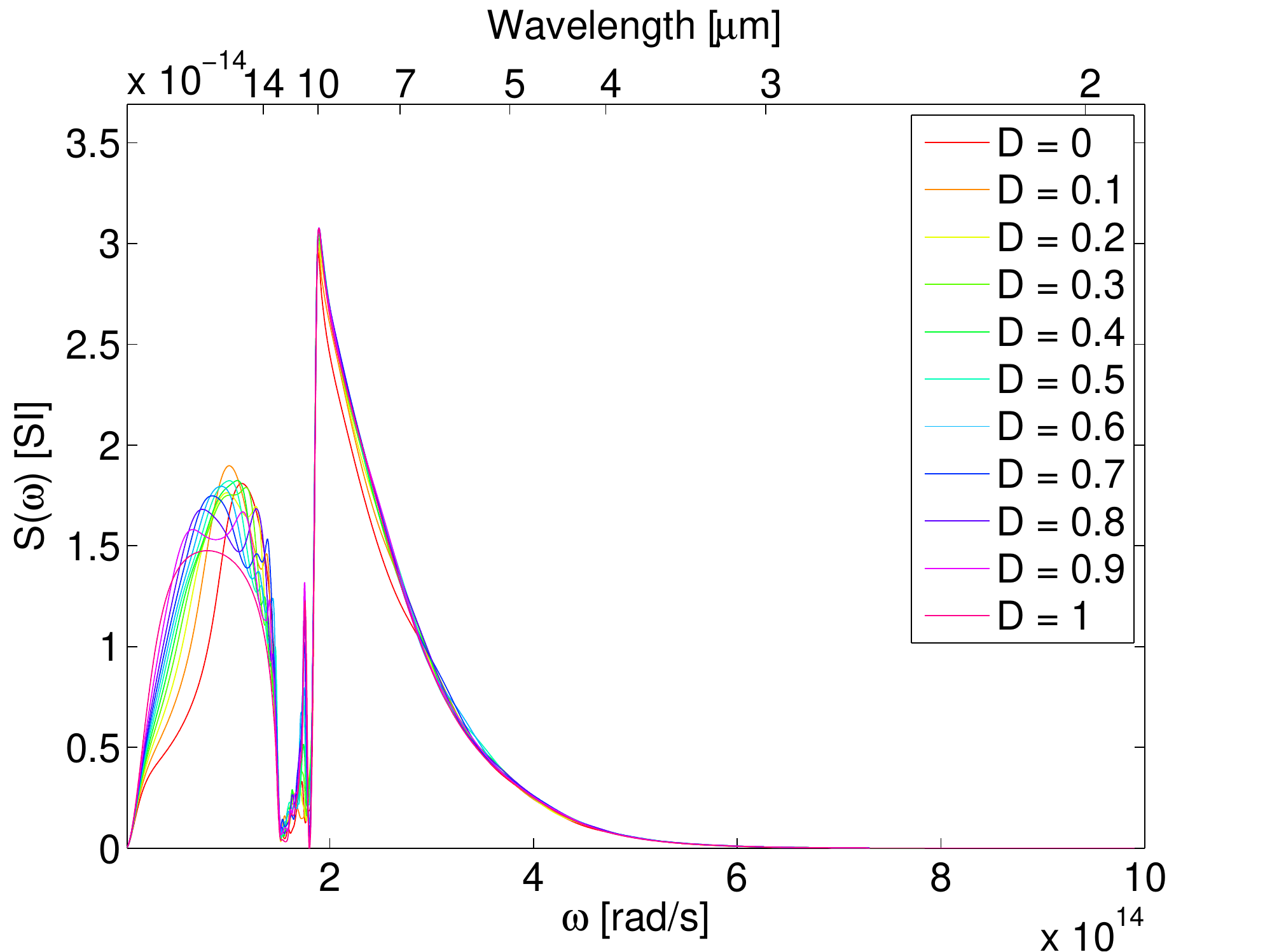}\includegraphics[scale=0.35]{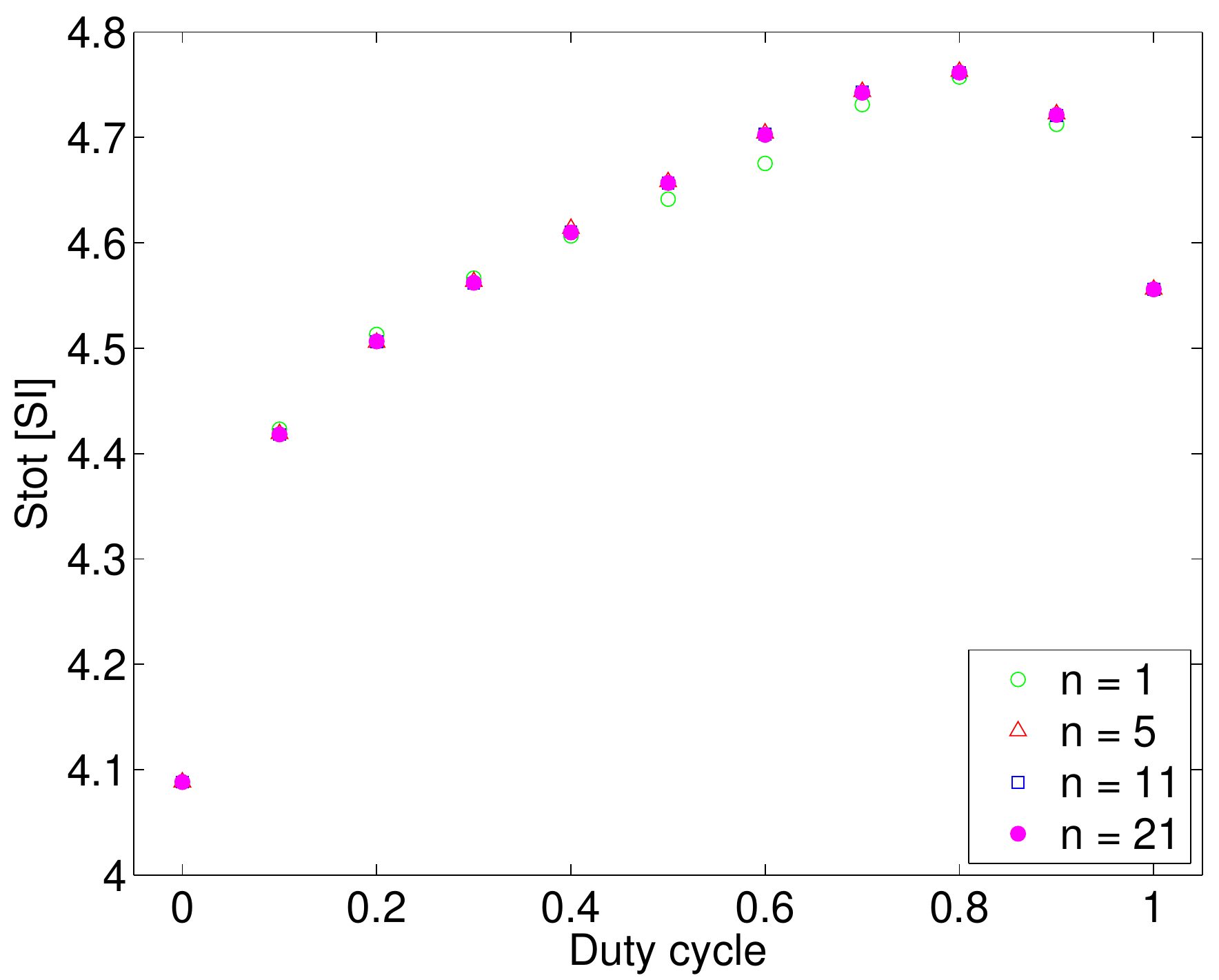}

\label{d_10um_Per_1um_fig}}\\
\subfloat[]{\includegraphics[scale=0.35]{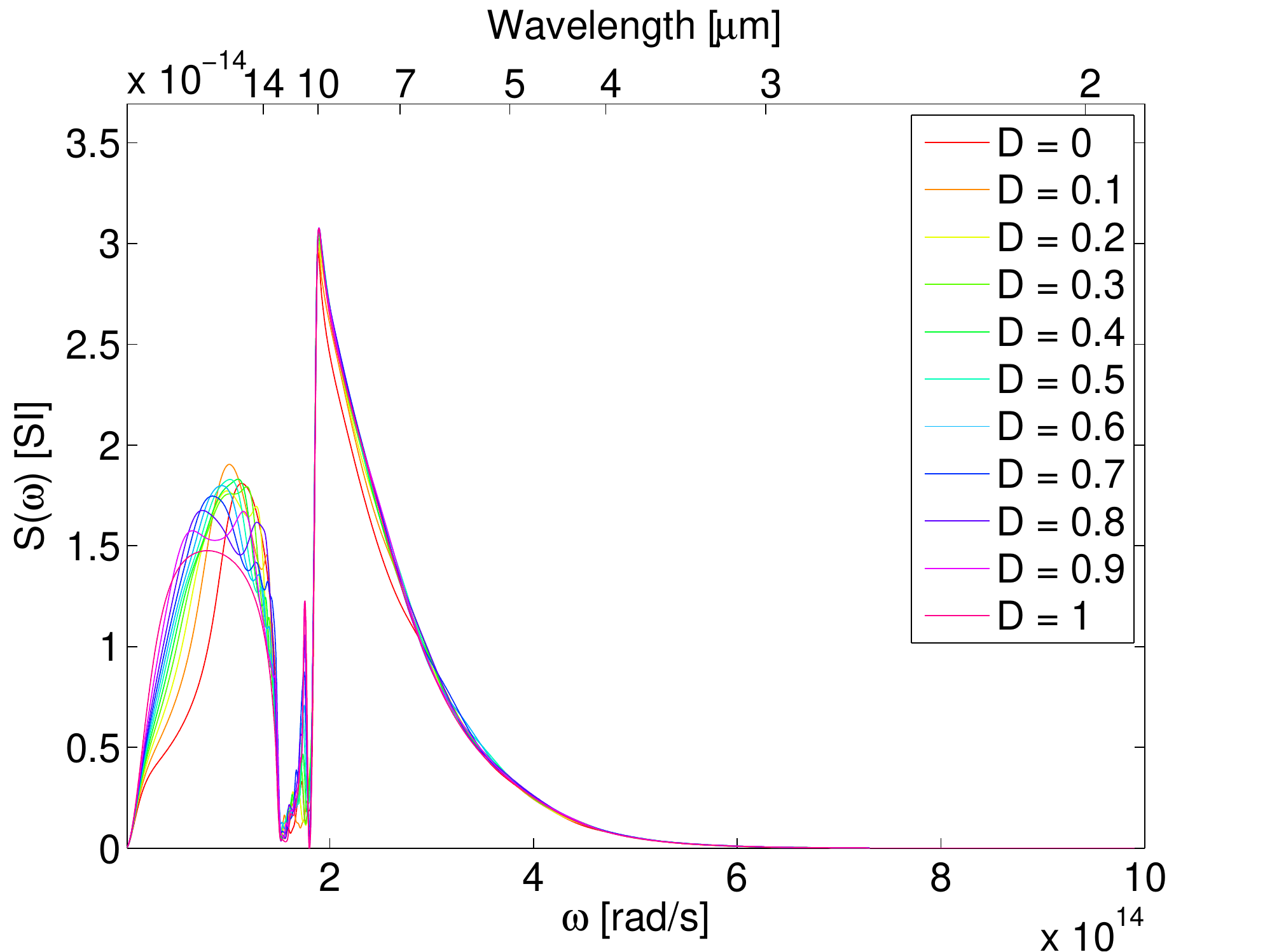}\includegraphics[scale=0.35]{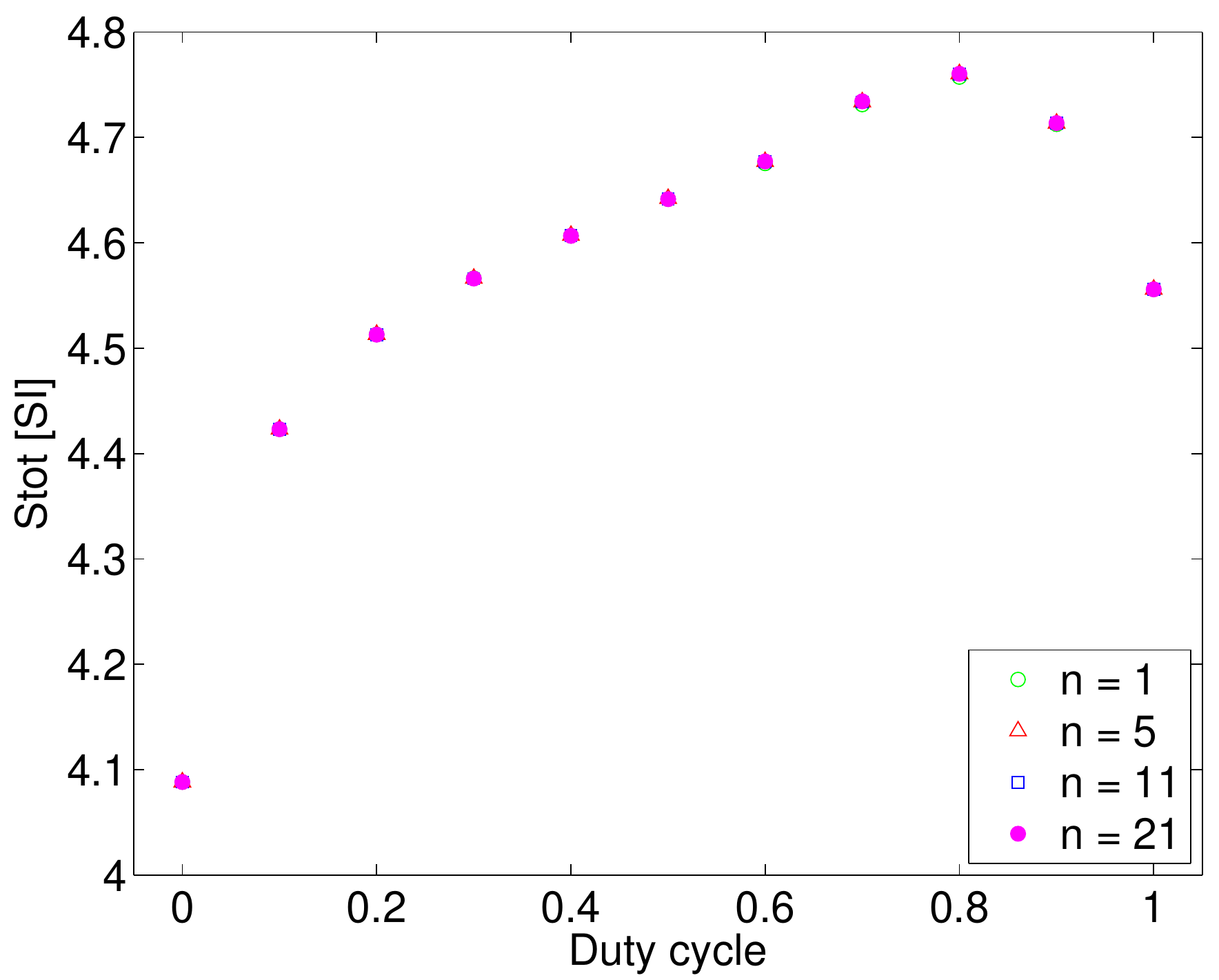}

\label{d_10um_Per_0.1um_fig}}\\

\par\end{centering}

\caption{Spectral contributions to the thermal capacitance and total thermal
capacitance for the structure shown in Fig. \ref{grating_fig} with
d = 10$\mu$m and different values of duty cycle with the periodicity
of (a) Per = 10$\mu$m (b) Per = 1$\mu$m (c) Per = 0.1$\mu$m}
\label{d_10um_fig}
\end{figure*}

\begin{figure*}
\begin{centering}
\subfloat[]{\includegraphics[scale=0.35]{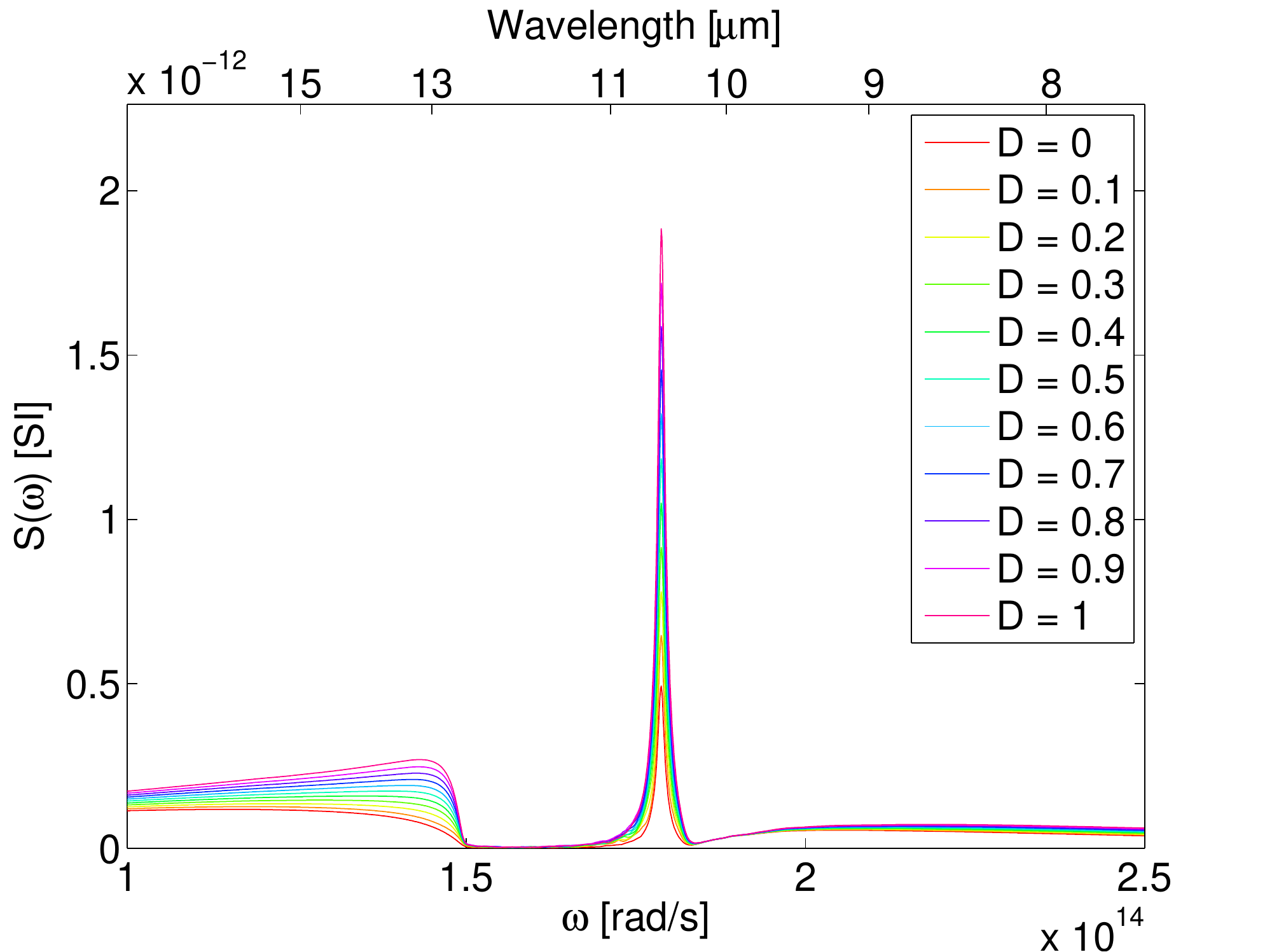}\includegraphics[scale=0.35]{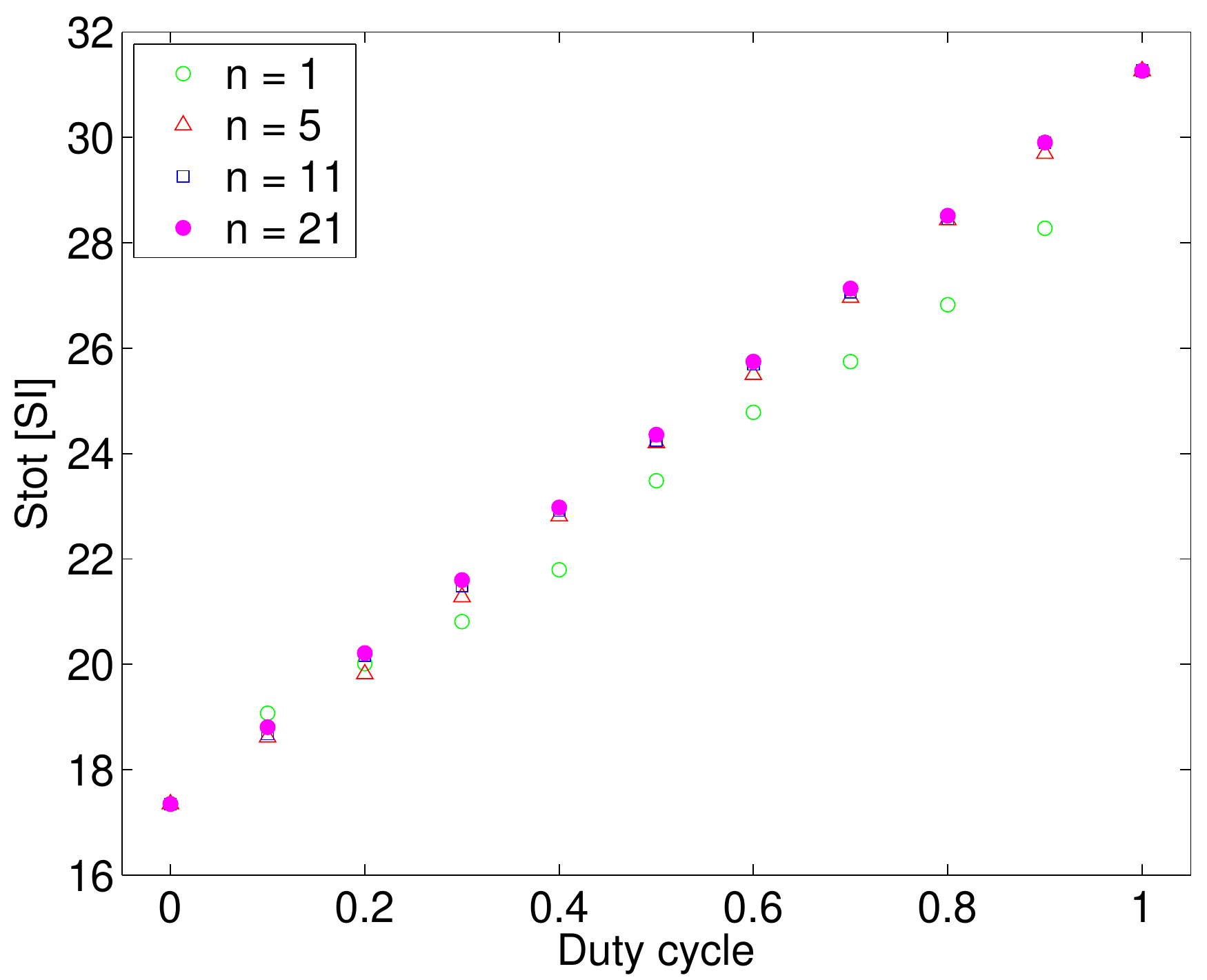}\label{d_1um_Per_10um_fig}}\\
\subfloat[]{\includegraphics[scale=0.35]{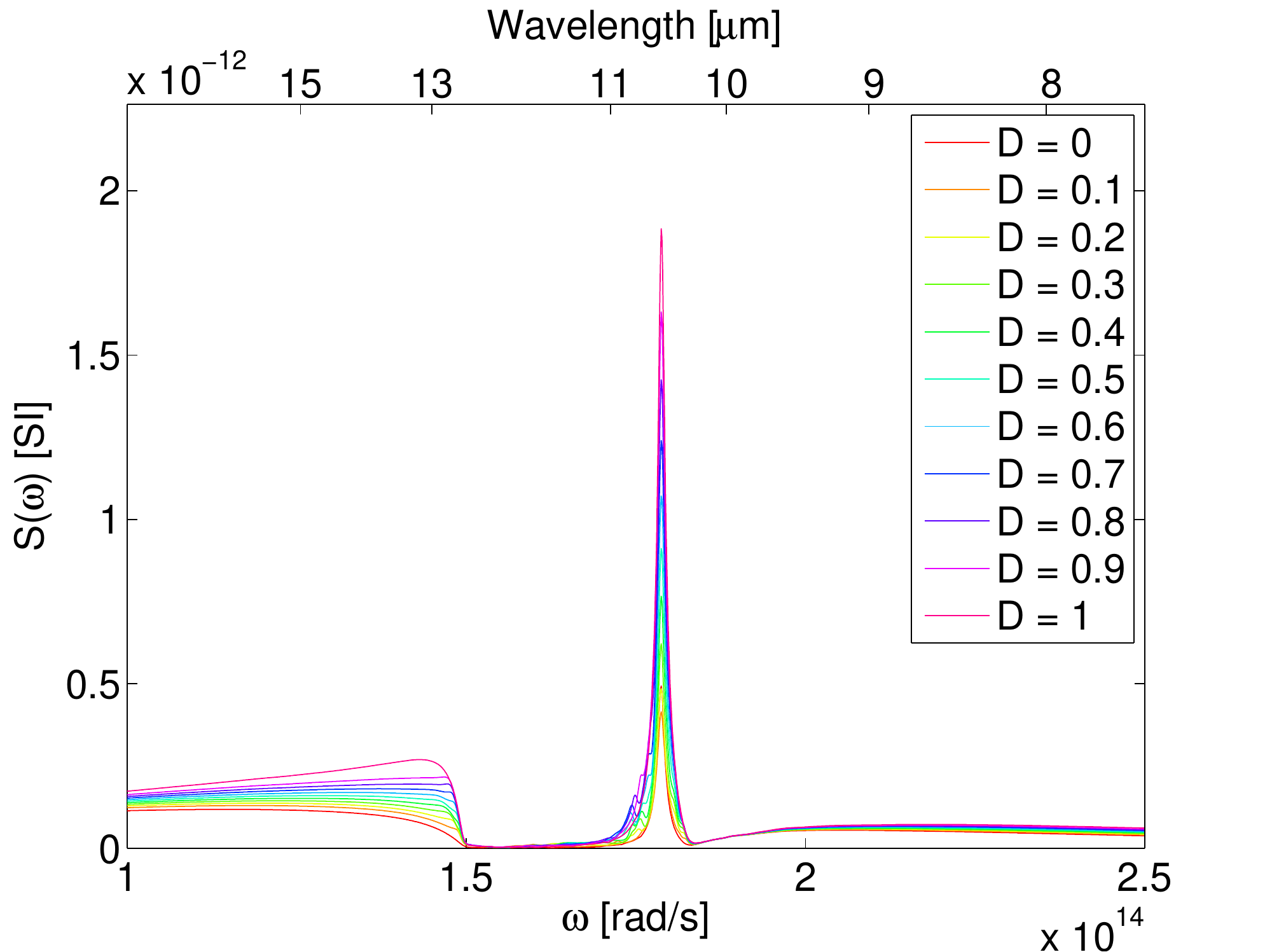}\includegraphics[scale=0.35]{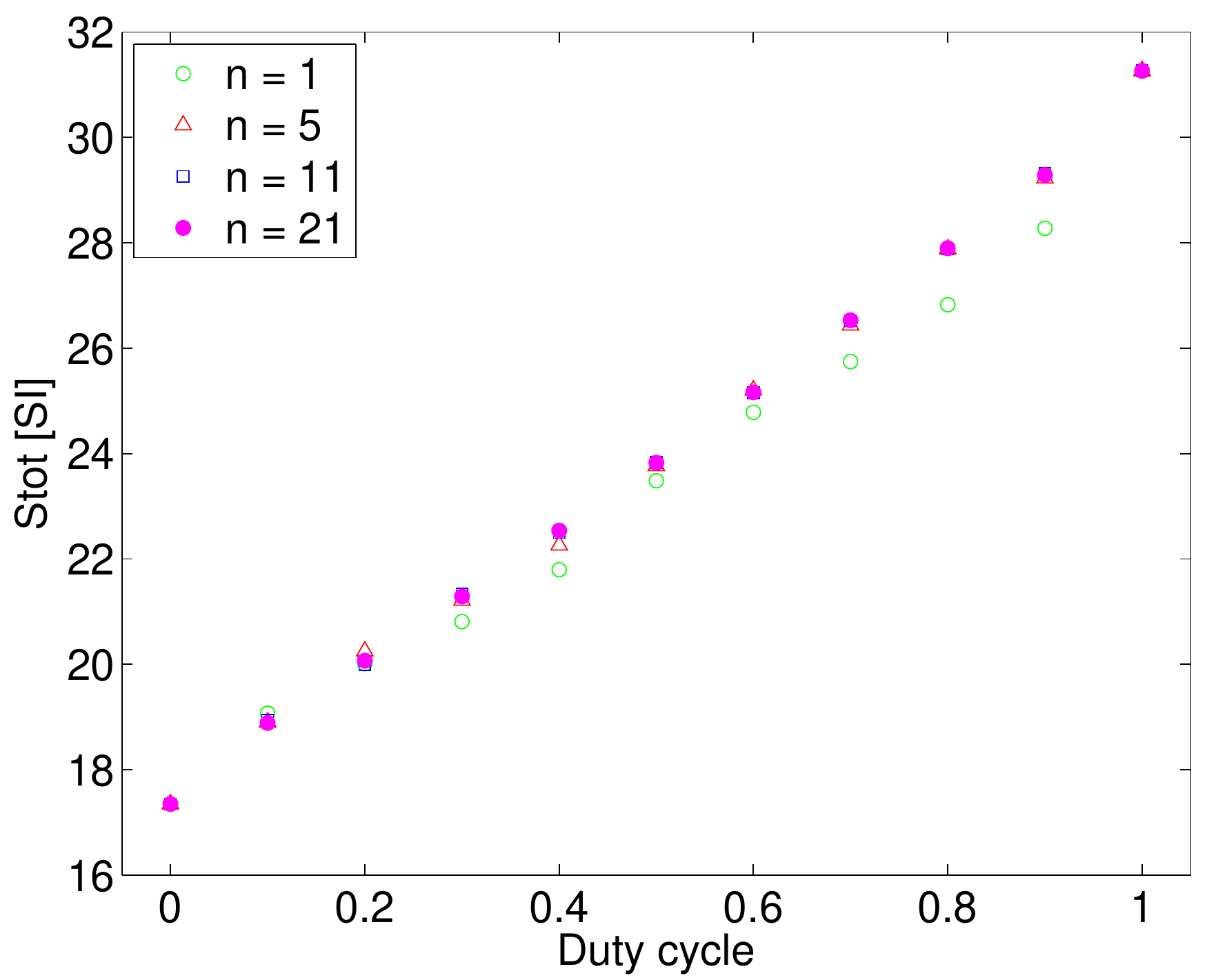}

\label{d_1um_Per_1um_fig}}\\
\subfloat[]{\includegraphics[scale=0.35]{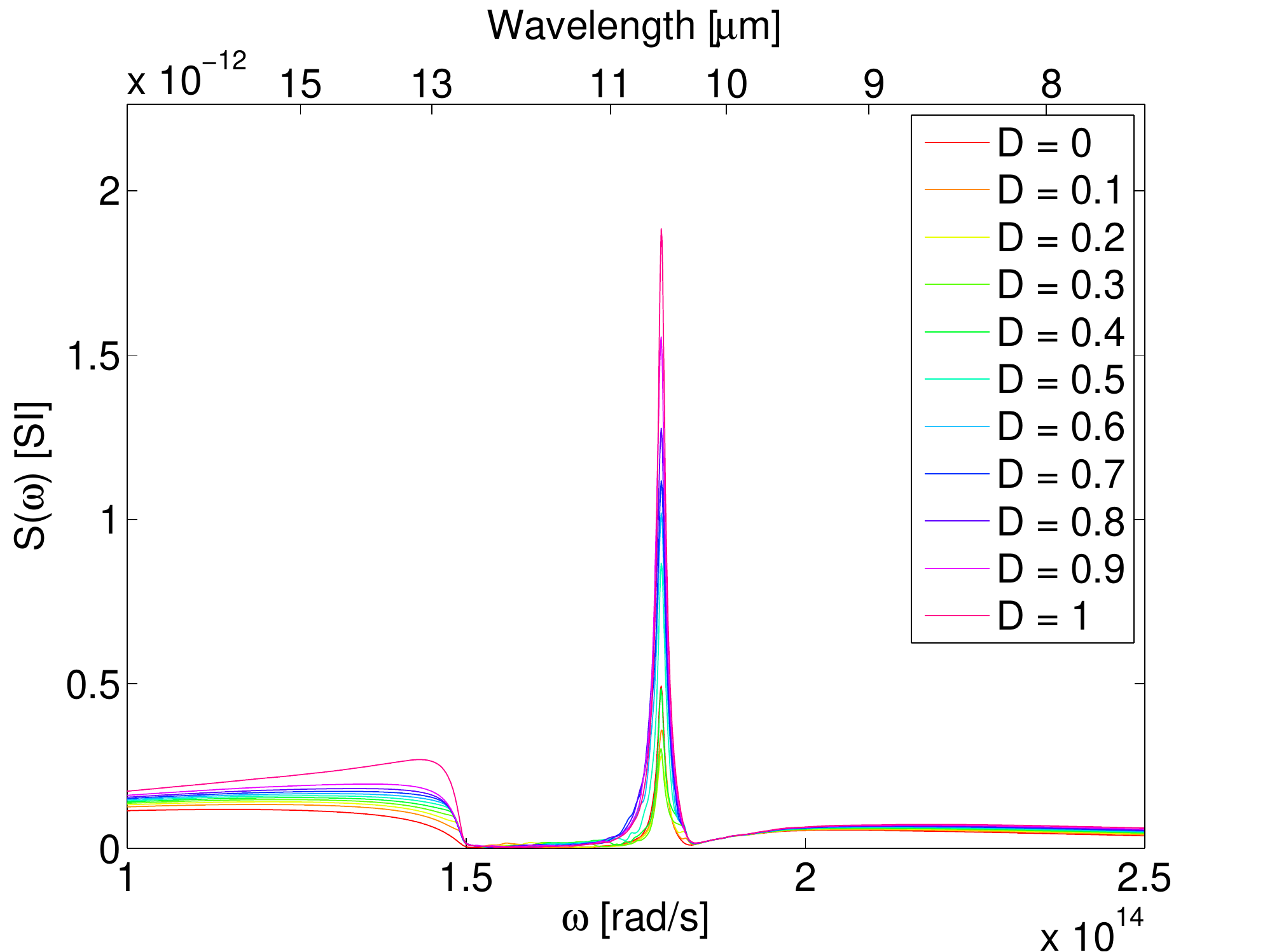}\includegraphics[scale=0.35]{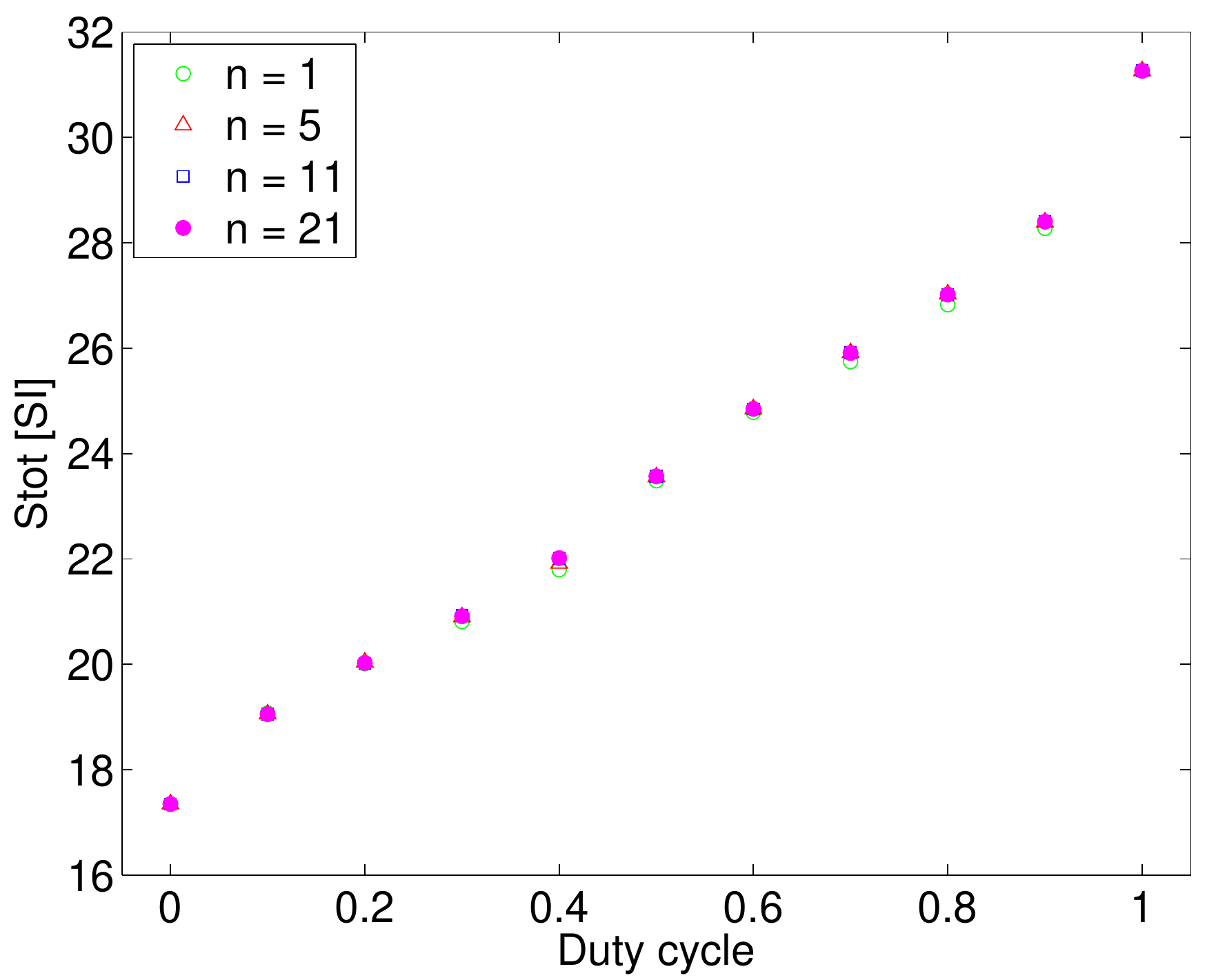}

\label{d_1um_Per_0.1um_fig}}\\

\par\end{centering}

\caption{Spectral contributions to the thermal capacitance and total thermal
capacitance for the structure shown in Fig. \ref{grating_fig} with
d = 1$\mu$m and different values of duty cycle with the periodicity
of (a) Per = 10$\mu$m (b) Per = 1$\mu$m (c) Per = 0.1$\mu$m}
\label{d_1um_fig}
\end{figure*}

\begin{figure*}
\begin{centering}
\subfloat[]{\includegraphics[scale=0.35]{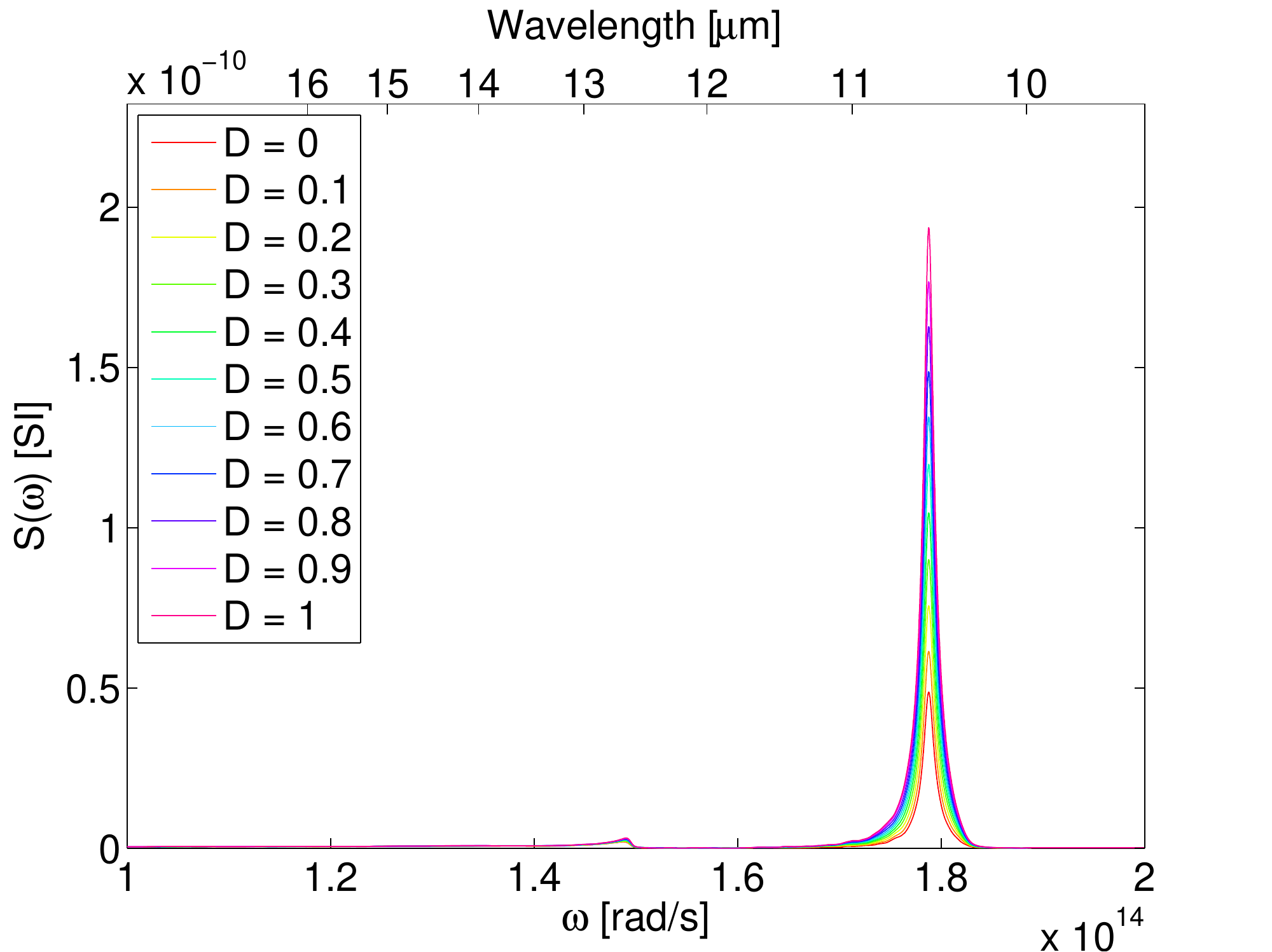}\includegraphics[scale=0.35]{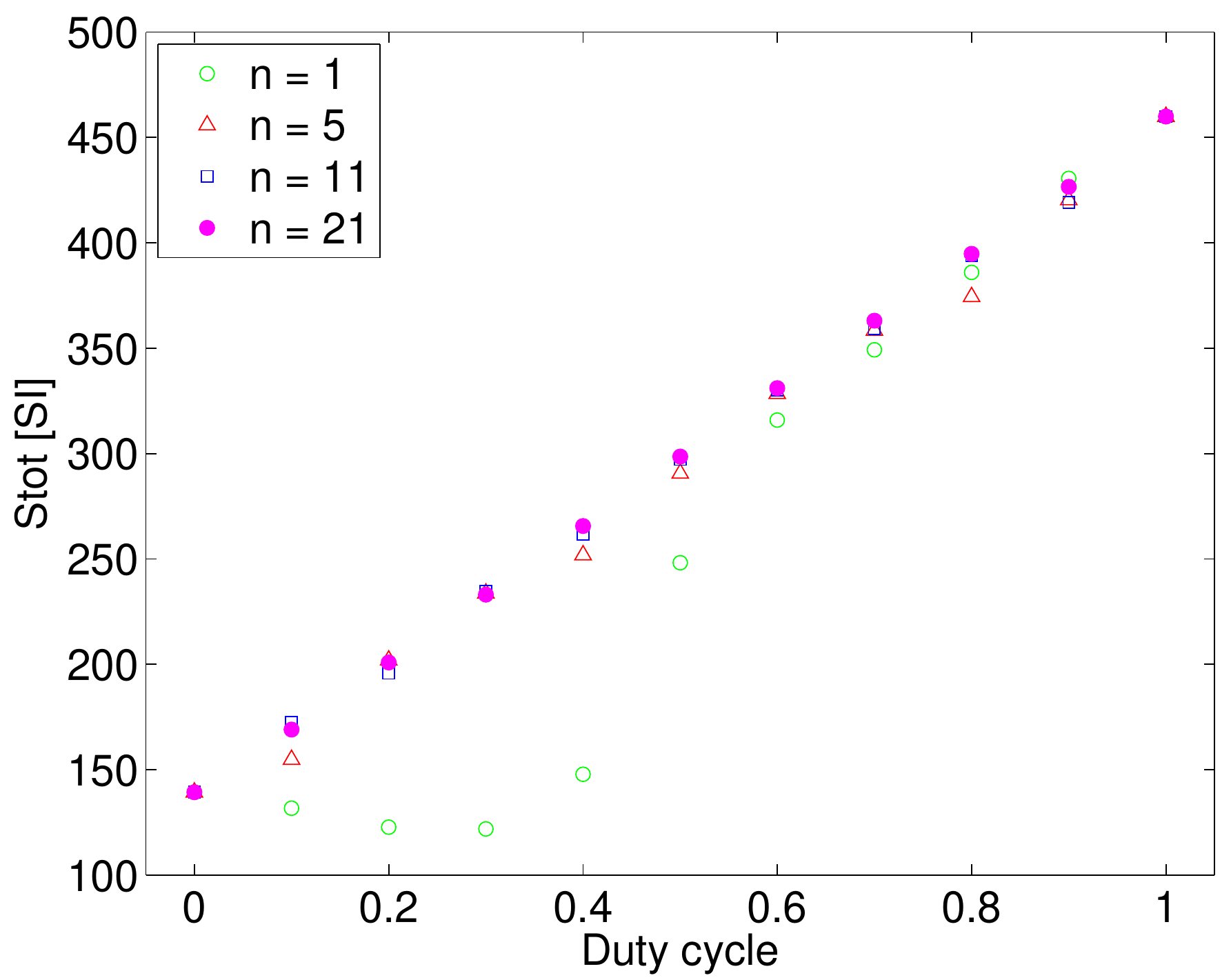}\label{d_0.1um_Per_10um_fig}}\\
\subfloat[]{\includegraphics[scale=0.35]{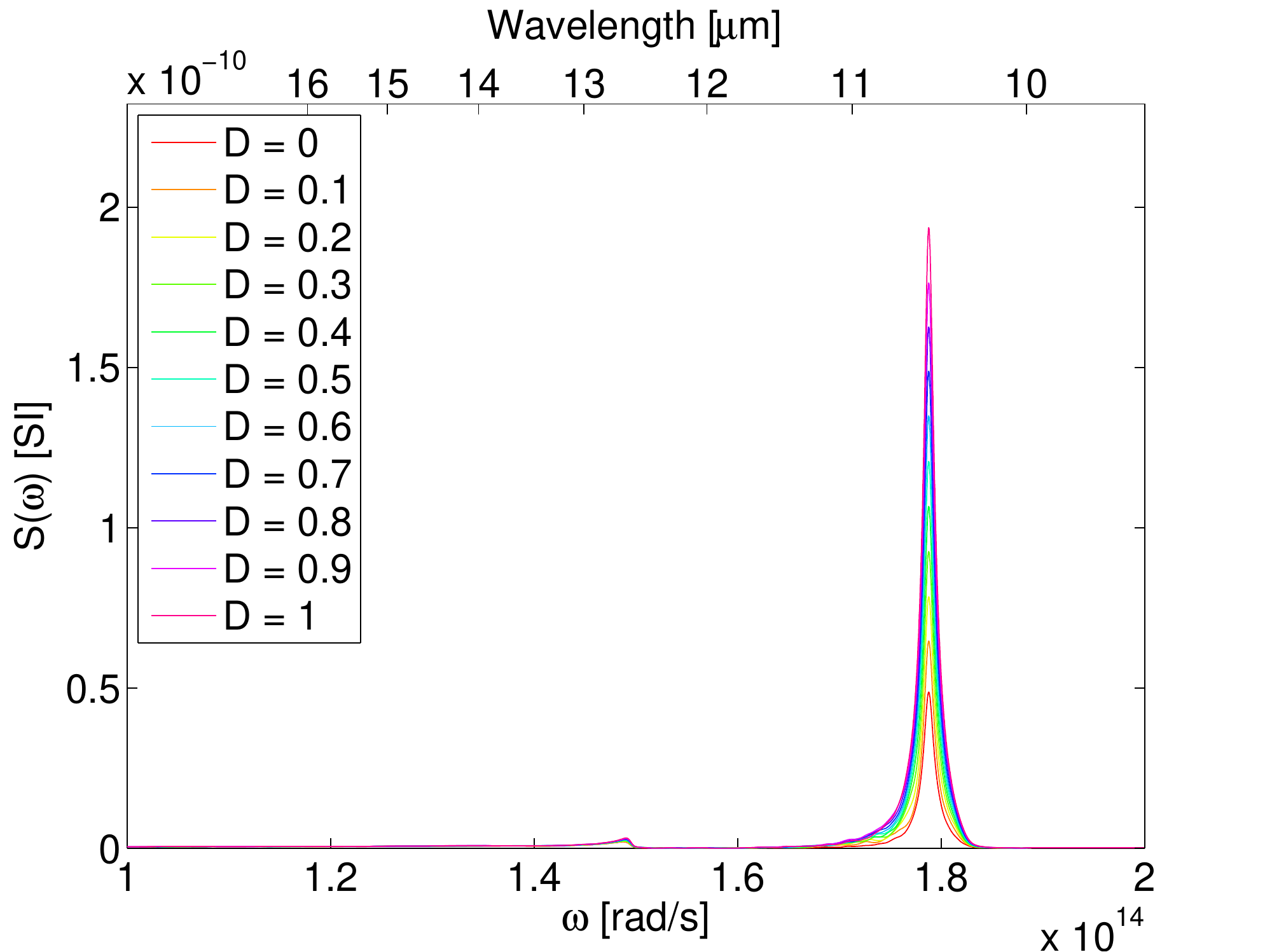}\includegraphics[scale=0.35]{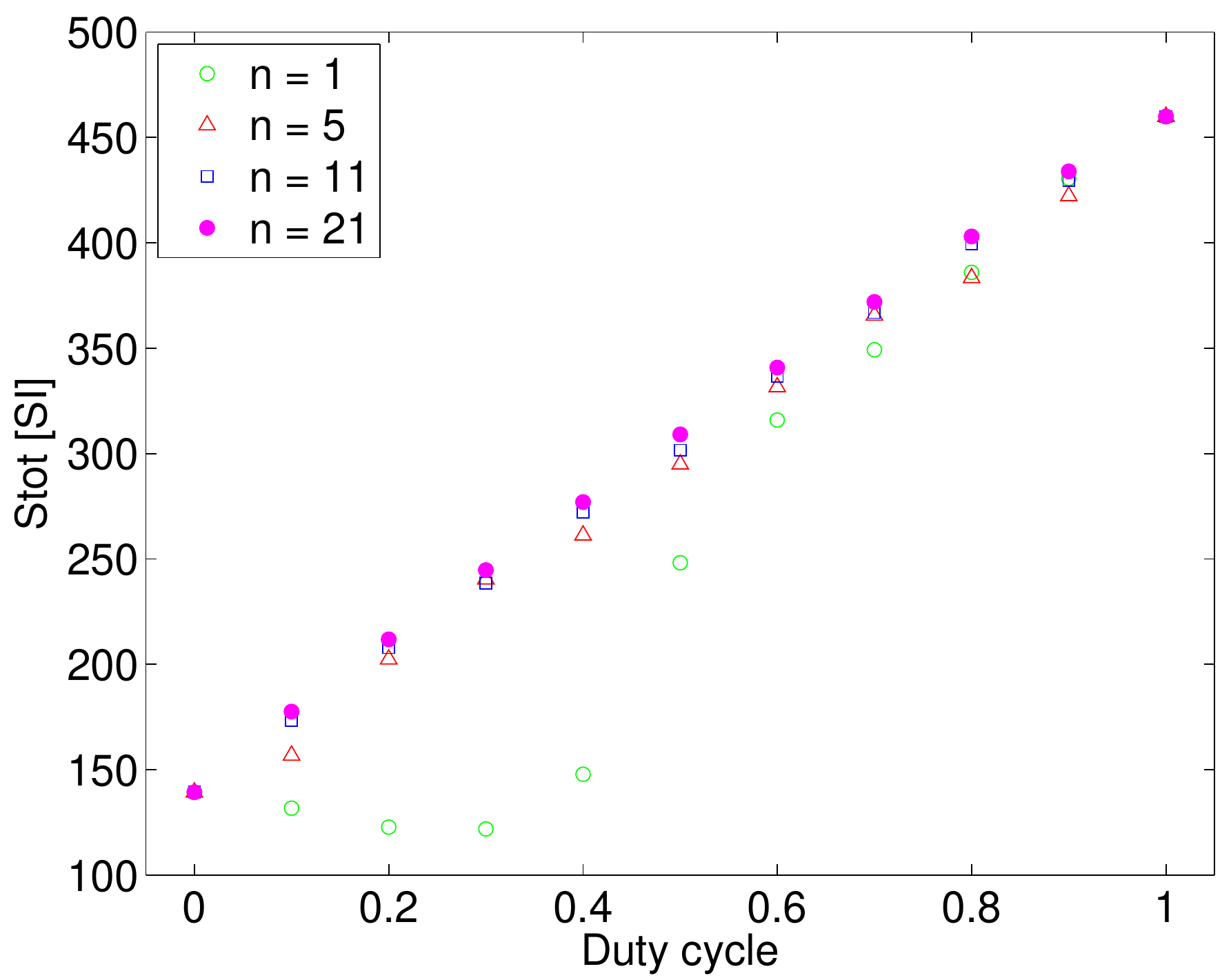}

\label{d_0.1um_Per_1um_fig}}\\
\subfloat[]{\includegraphics[scale=0.35]{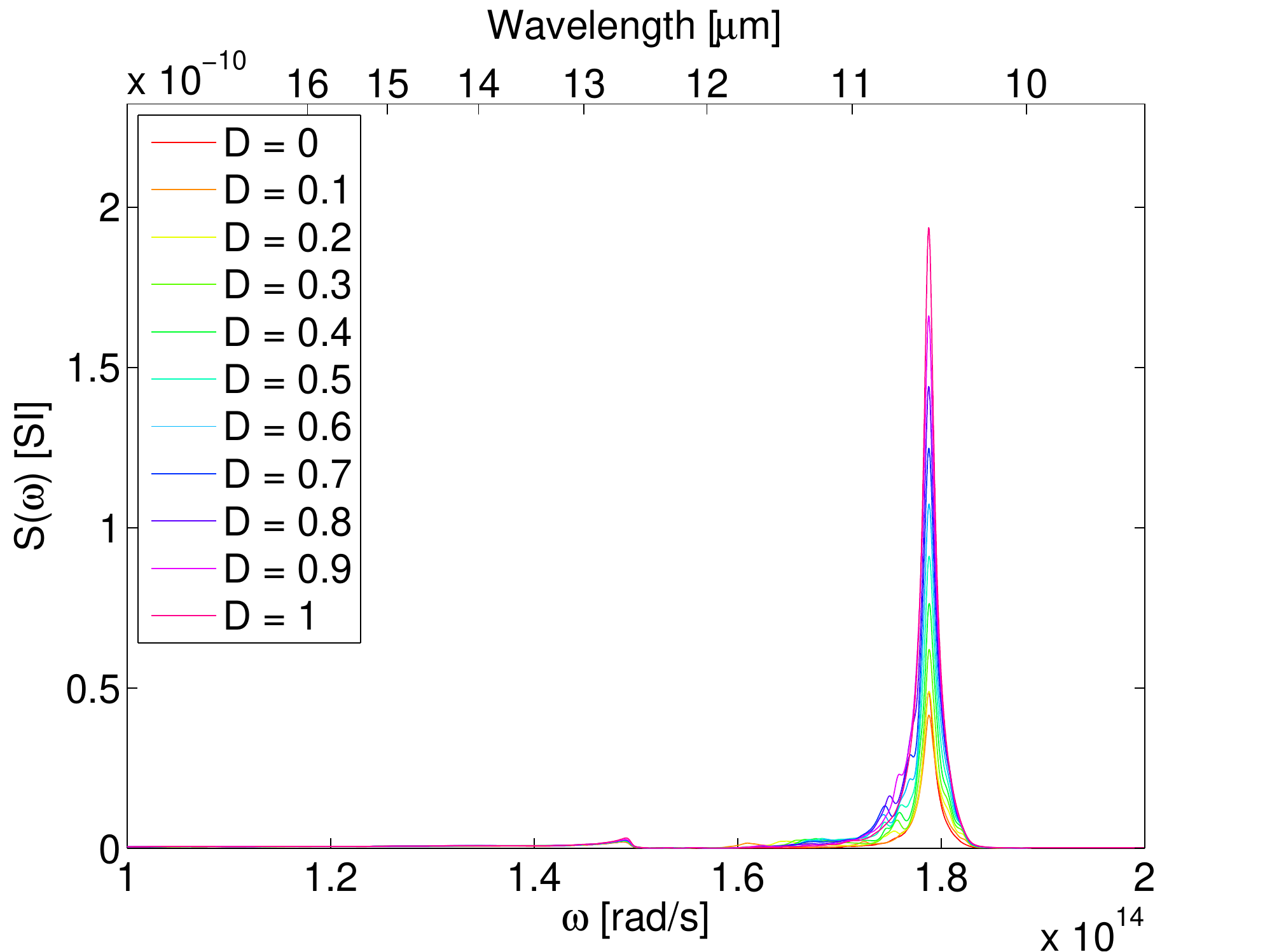}\includegraphics[scale=0.35]{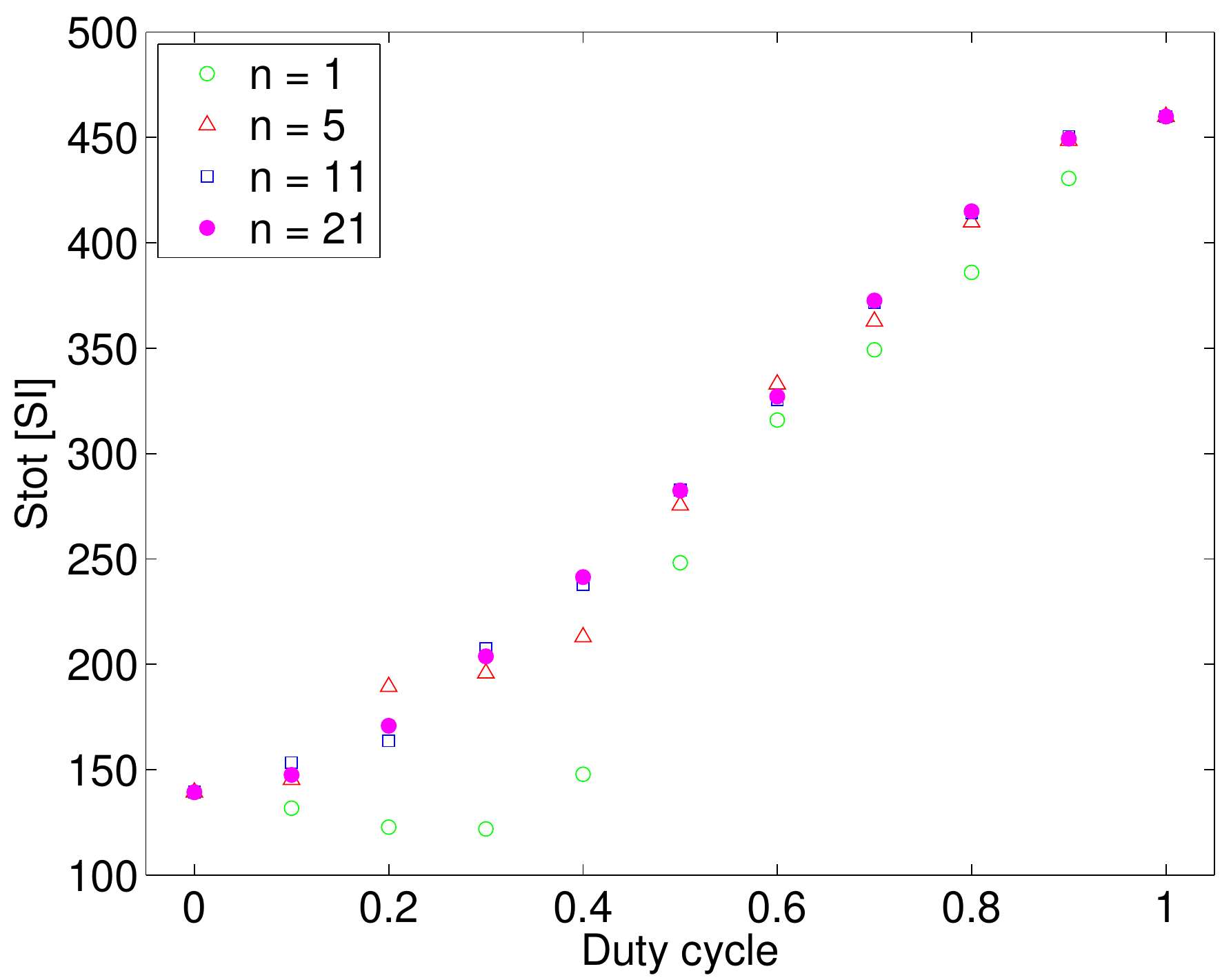}

\label{d_0.1um_Per_0.1um_fig}}\\

\par\end{centering}

\caption{Spectral contributions to the thermal capacitance and total thermal
capacitance for the structure shown in Fig. \ref{grating_fig} with
d = 0.1$\mu$m and different values of duty cycle with the periodicity
of (a) Per = 10$\mu$m (b) Per = 1$\mu$m (c) Per = 0.1$\mu$m}
\label{d_0.1_fig}
\end{figure*}

The results of calculation for the structure shown in Fig. \ref{grating_fig}
at different values of periodicities and distances are summarized
in Figs. \ref{d_10um_fig}, \ref{d_1um_fig}, and \ref{d_0.1_fig}
respectively for $d=10\mu m$, $d=1\mu m$, and $d=0.1\mu m$. As
the figures show, convergence is achieved with incorporation of 21
harmonics in all cases. The total thermal capacitance in most cases
is monotonically increasing with increasing duty cycle. This comes
from the fact that gratings with higher duty cycles feature more SiC
material that is located near the adjacent SiC slab. This then naturally
facilitates higher evanescent coupling. However, for the case of $d=10\mu m$,
a peak in thermal transfer is achieved for a value of duty cycle which
is neither 0 or 1. Noting the fact that the resonance wavelength of
the SiC is around $10\mu m$, we expect Mie resonances of the beam
to become important in that case and give rise to the highest thermal
capacitance for the non-unity duty cycle. 

One important fact that can be derived from the obtained results is
that, as the periodicity decreases, the result obtained with using
just one harmonic becomes more accurate. This is to be expected since
in the case of incorporating just one harmonic, our method reproduces
results obtained by the effective medium theory which becomes more
and more accurate in subwavelength regime (compared with the resonance
wavelength).

\begin{figure}
\begin{centering}
\subfloat[]{\begin{centering}
\includegraphics[scale=0.3]{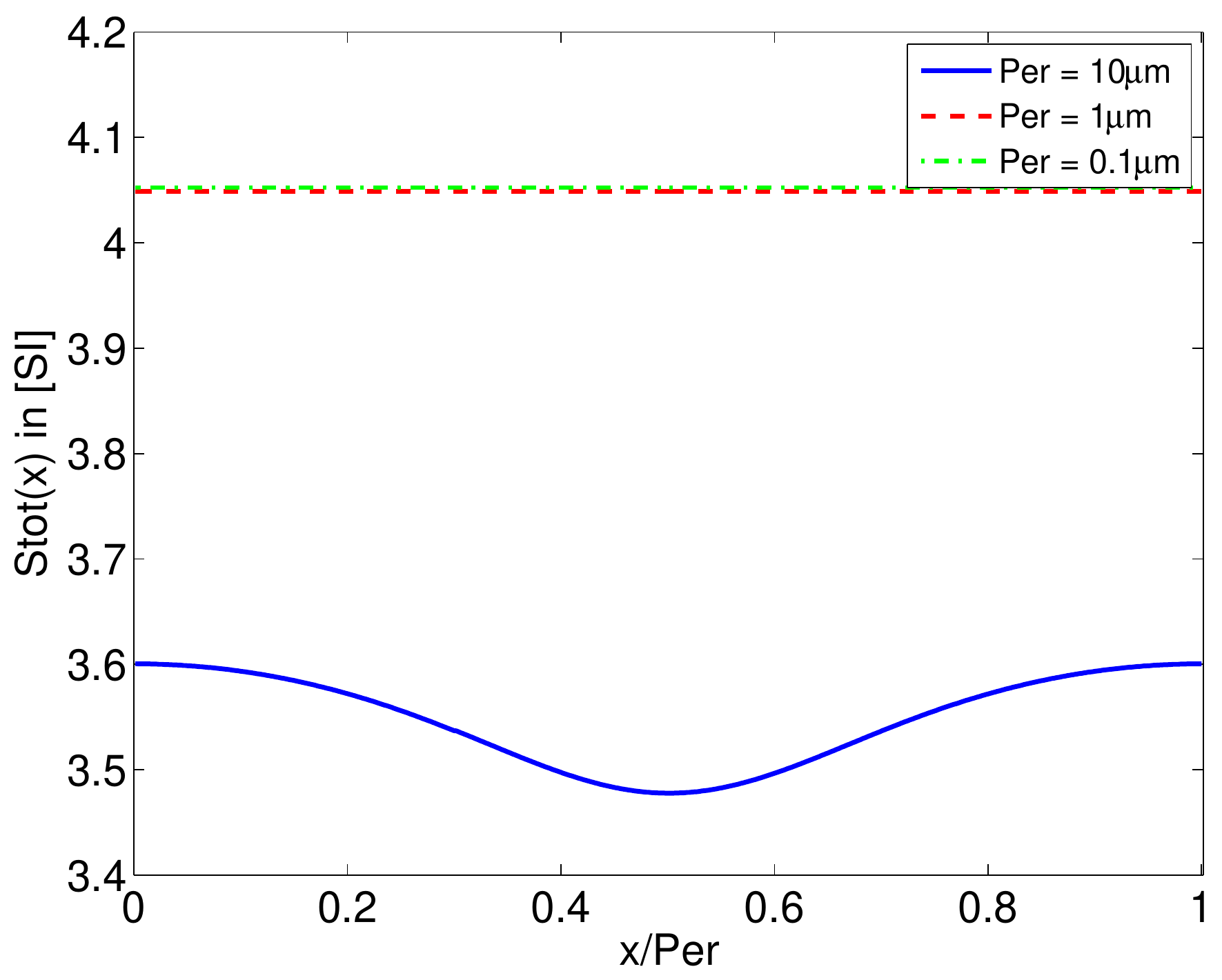}
\par\end{centering}

}\\
\subfloat[]{\begin{centering}
\includegraphics[scale=0.3]{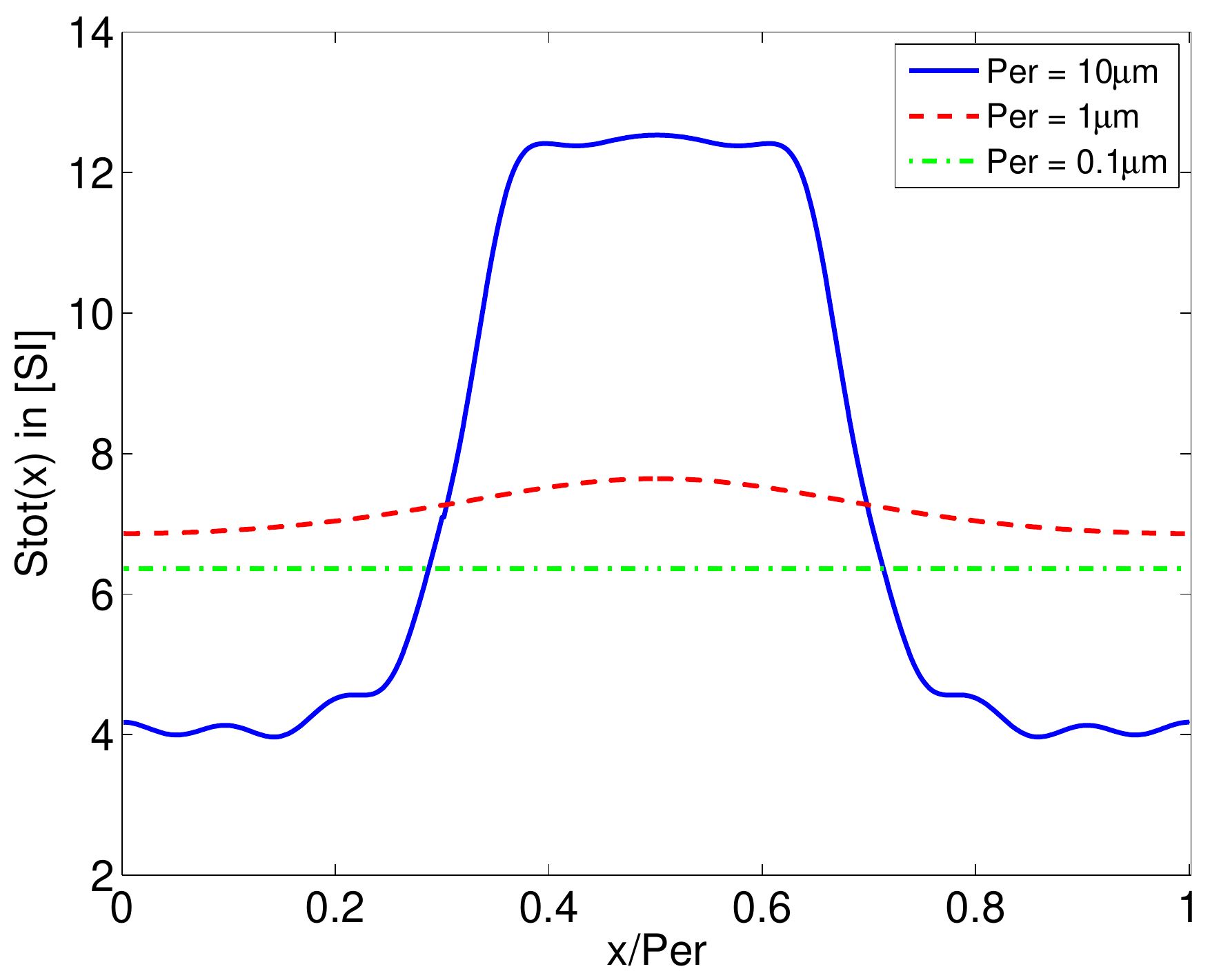}
\par\end{centering}

}\\
\subfloat[]{\centering{}\includegraphics[scale=0.3]{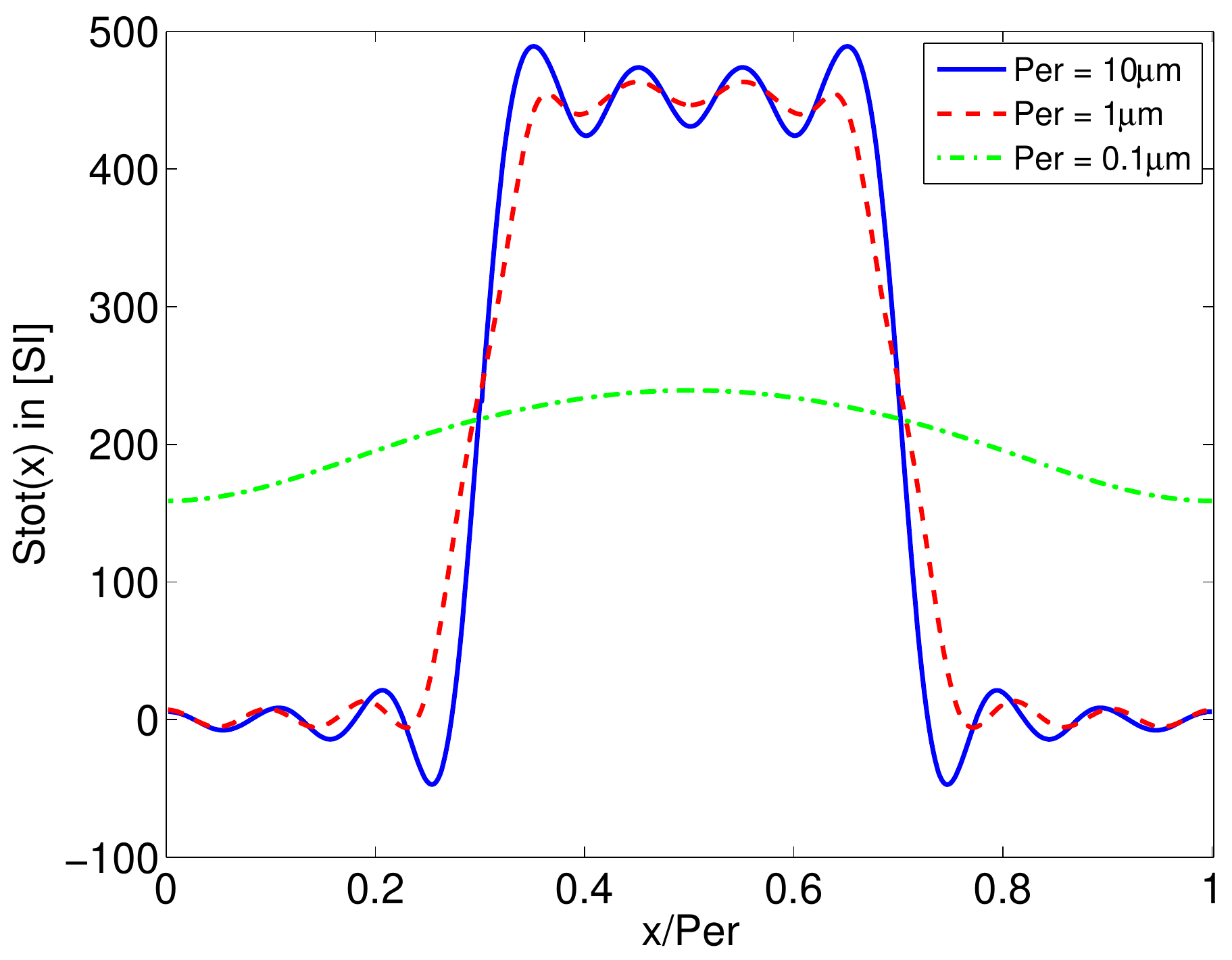}}
\par\end{centering}

\caption{Contributions to the thermal capacitance for the structure shown in
Fig. \ref{rods_fig} across a period for different values of periodicity,
and assuming a constant value of $D=0.4$, in the case of (a) $d=10\mu m$
(b) $d=1\mu m$ (c) $d=0.1\mu m$}

\label{newd_xvar_fig}
\end{figure}

This fact is most evident in Fig. \ref{d_xvar_fig}, which shows the
contribution of different points across the period to the total thermal
capacitance. In this figure the thermal capacitance is plotted as
a function of $x$ position across the period at $z=0$ plane for
different values of periodicity and distances. The duty cycle is assumed
to be the same value of $0.4$ in all cases. Note that since we have
translational symmetry in the y direction, there is no change in thermal
capacitance in that direction. This figure verifies the fact that
thermal capacitance in the limit of small periods, tends toward a
constant value across the period which is determined by effective
medium theory. On the other hand, this figure further demonstrates
the fact that for periodicities larger than some critical value, thermal
capacitance can be modeled as a superposition of two channels; a channel
with larger thermal capacitance which is due to parts of slabs that
are closer together and the other one with smaller thermal capacitance
which is due to the sections that are farther from each other. These
plots were obtained by incorporation of 21 harmonics.

\begin{figure*}
\begin{centering}
\subfloat[]{\includegraphics[scale=0.35]{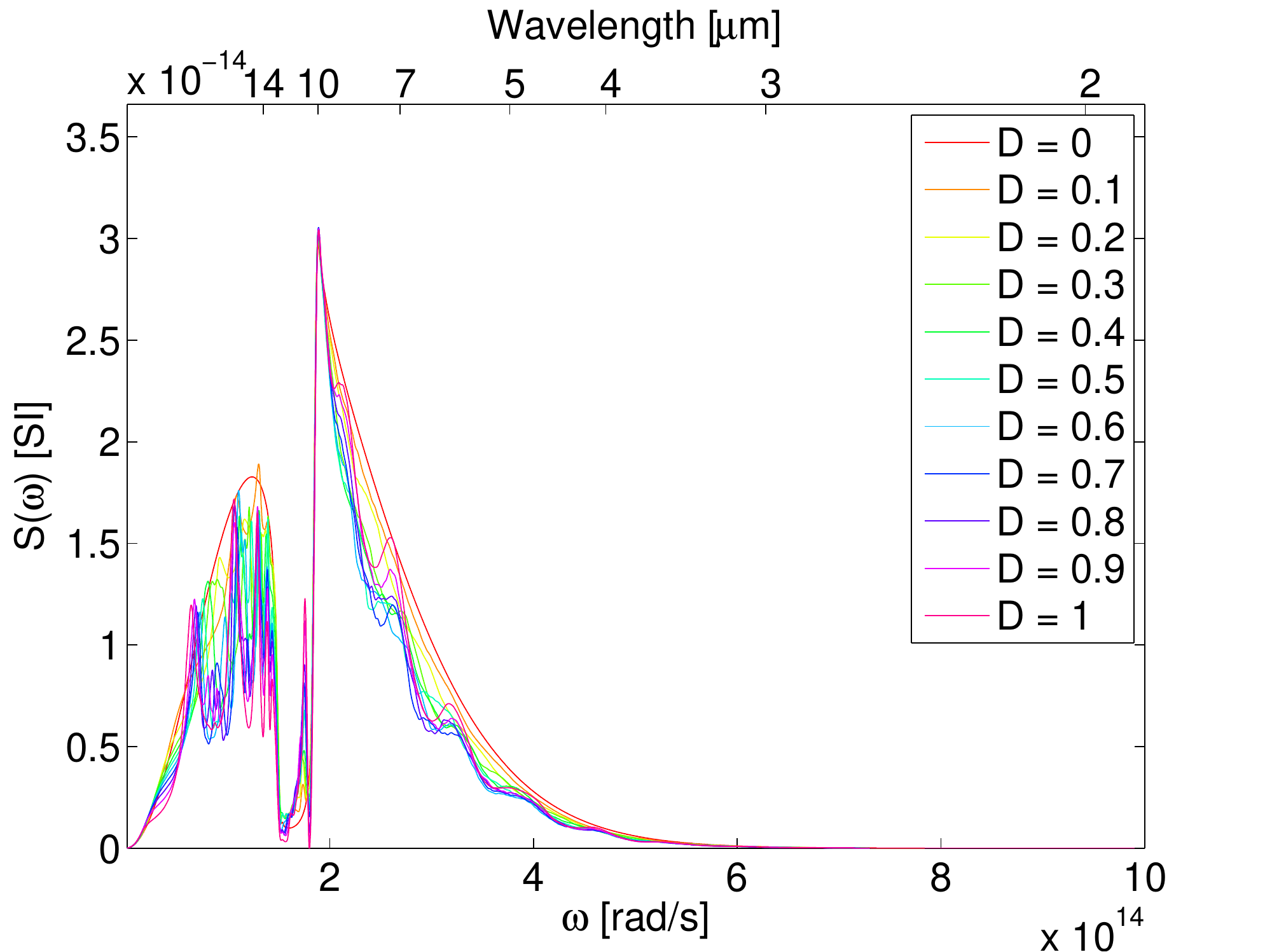}\includegraphics[scale=0.35]{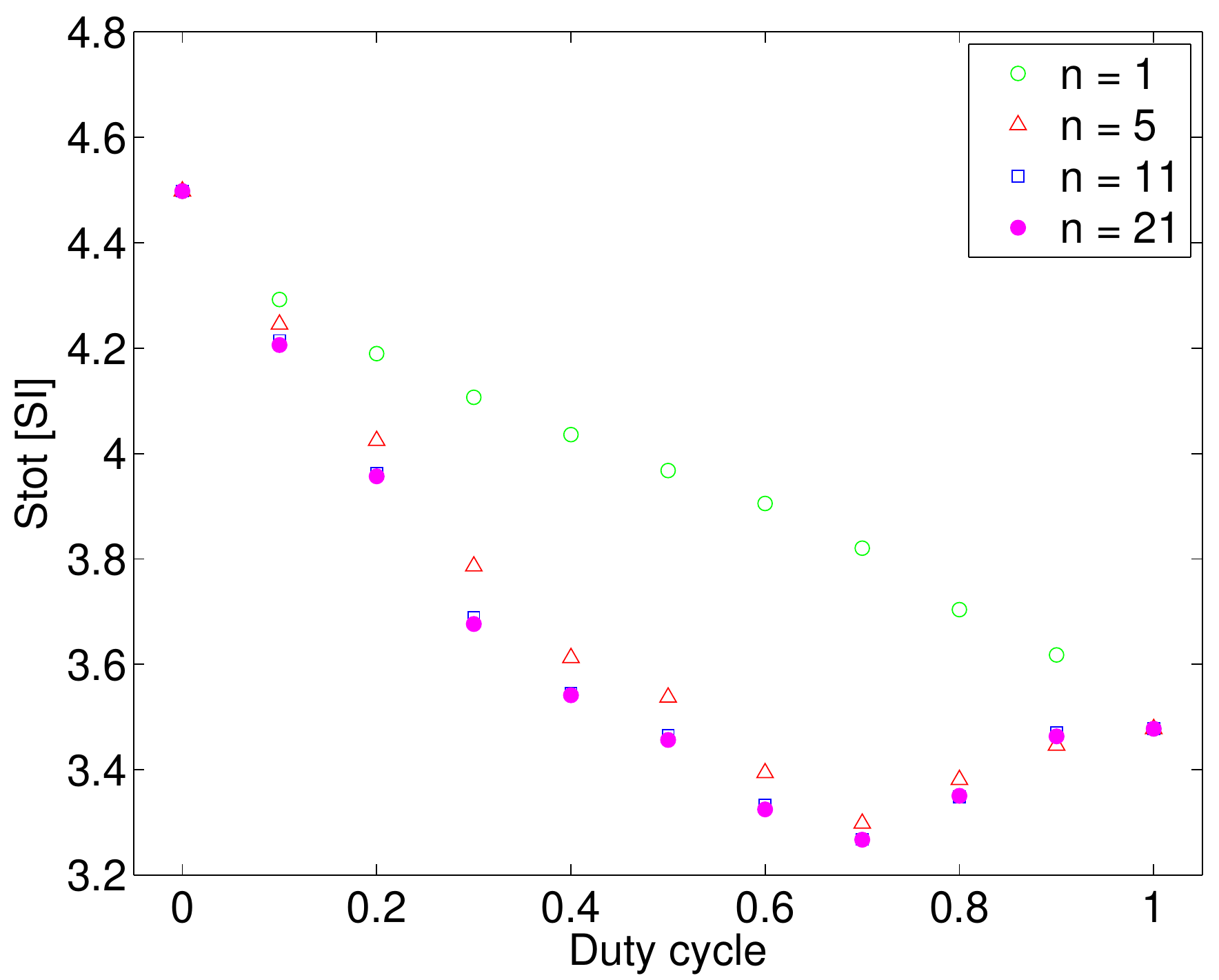}\label{newd_10um_Per_10um_fig}}\\
\subfloat[]{\includegraphics[scale=0.35]{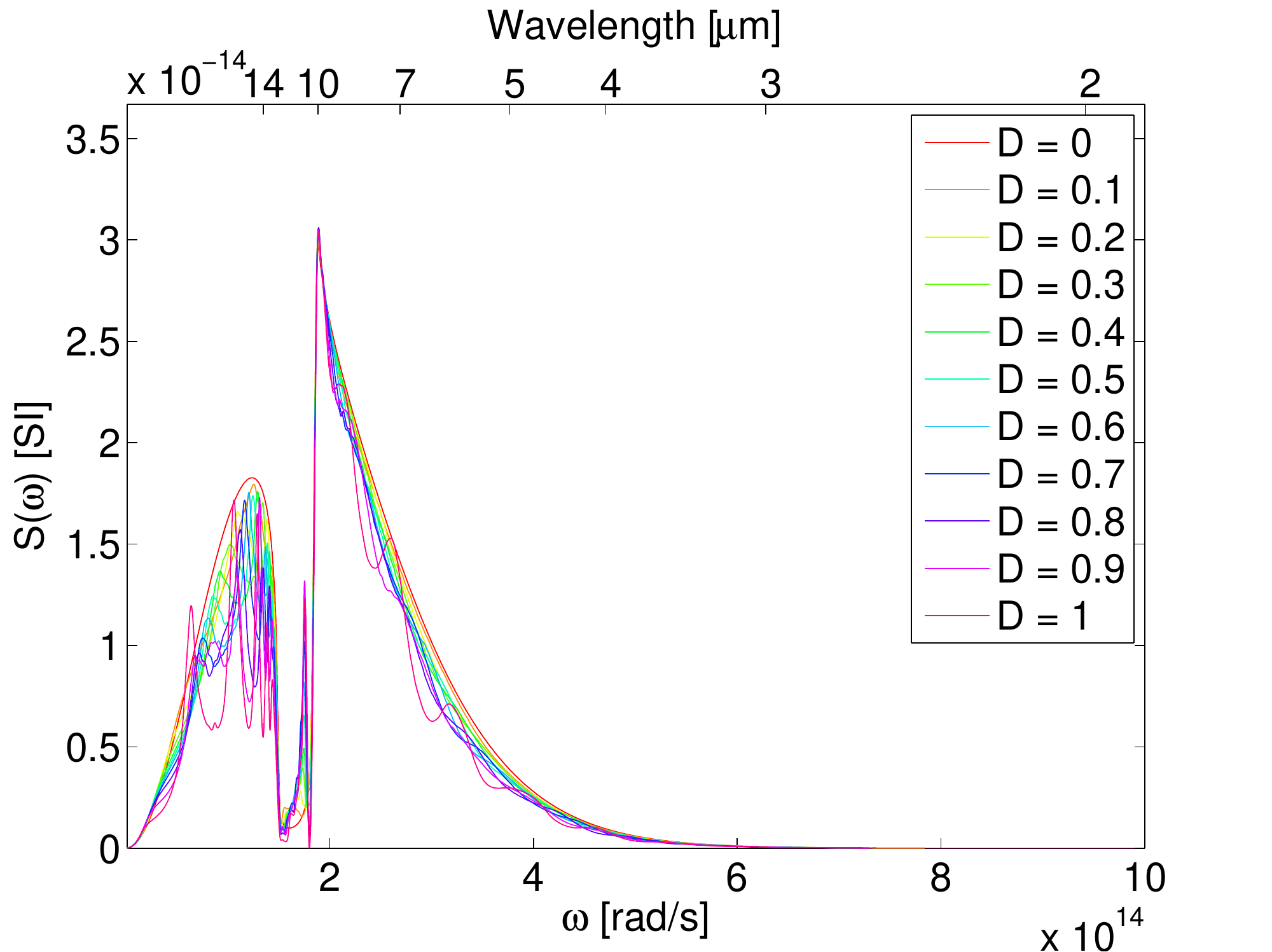}\includegraphics[scale=0.35]{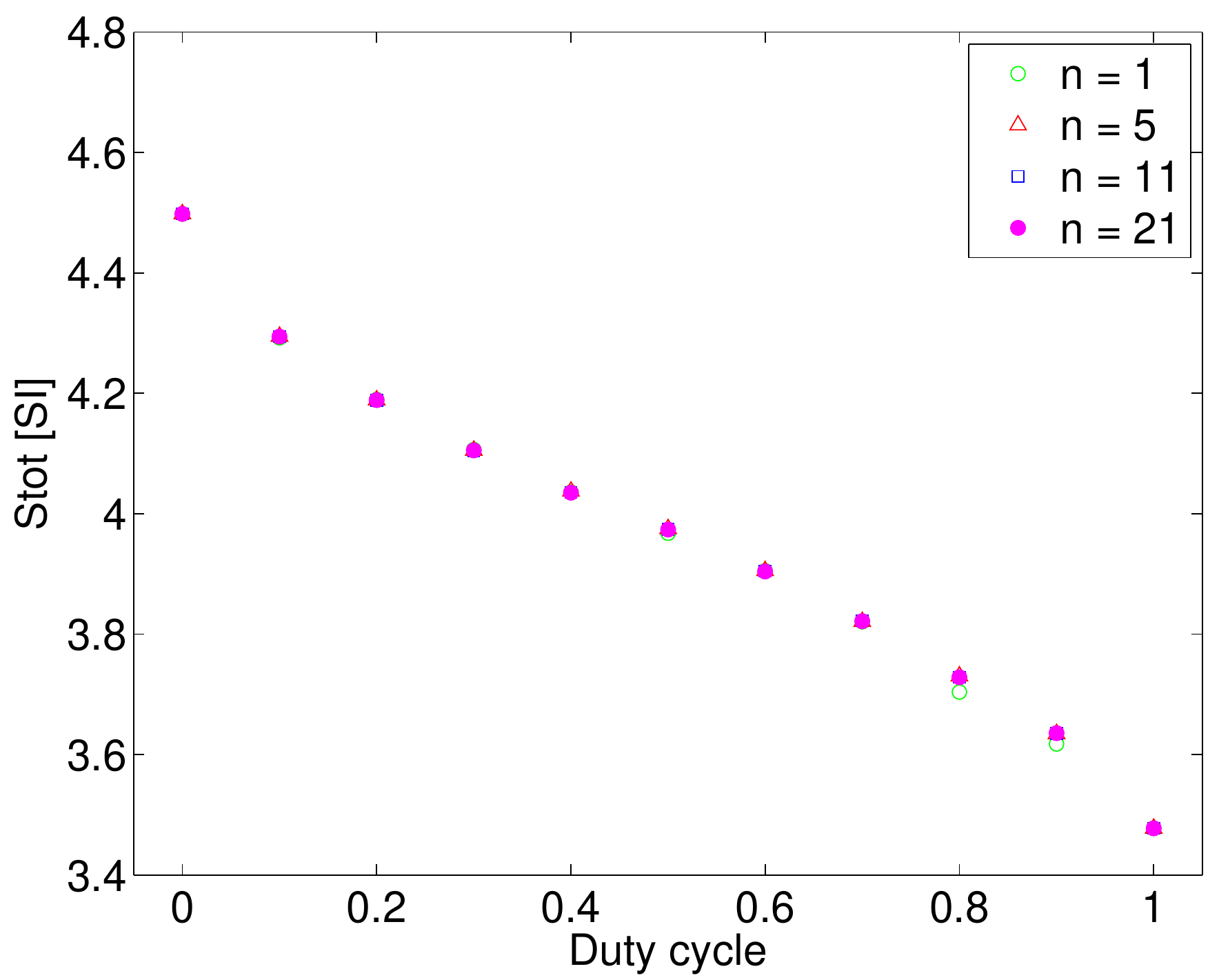}

\label{newd_10um_Per_1um_fig}}\\
\subfloat[]{\includegraphics[scale=0.35]{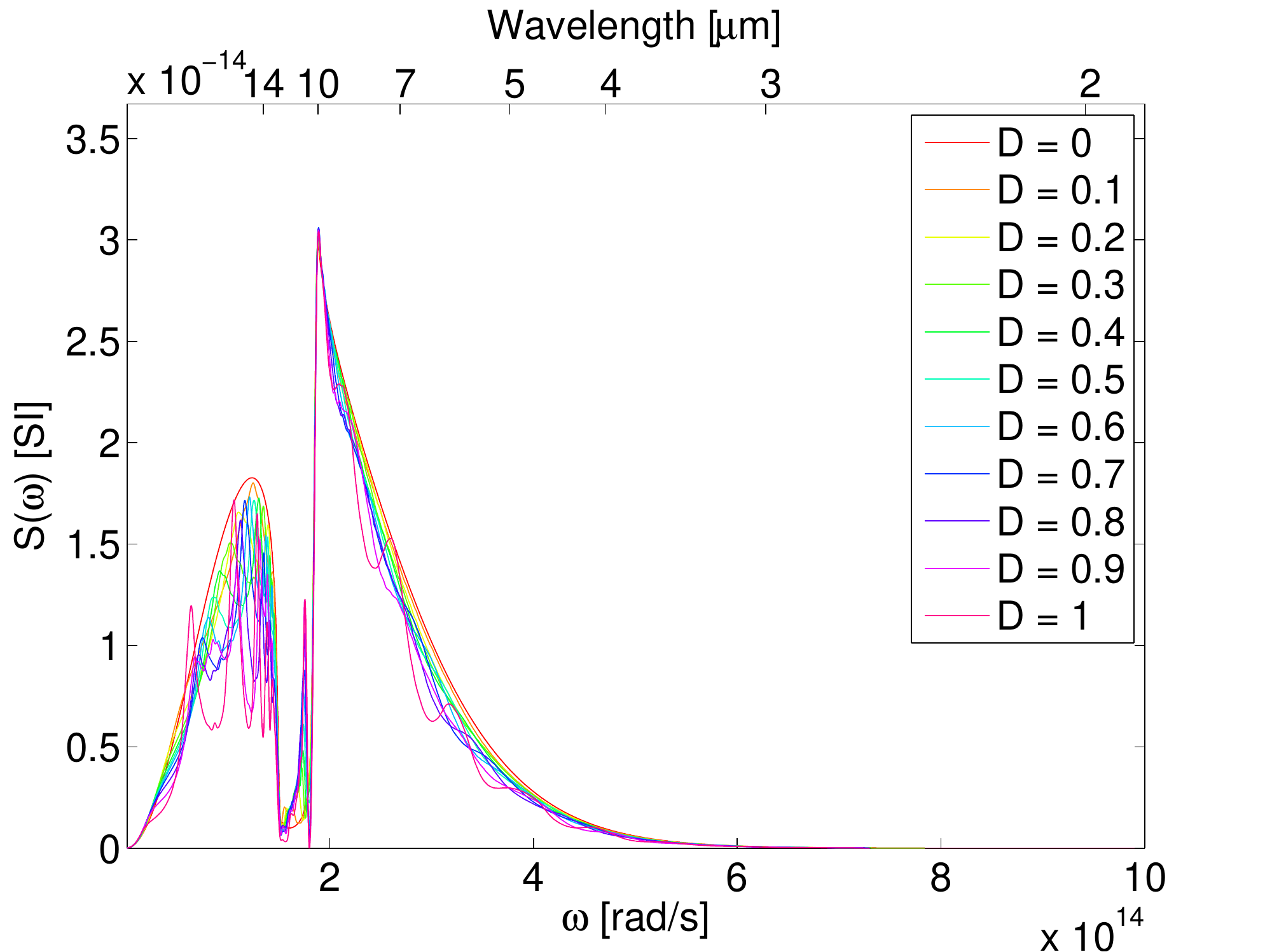}\includegraphics[scale=0.35]{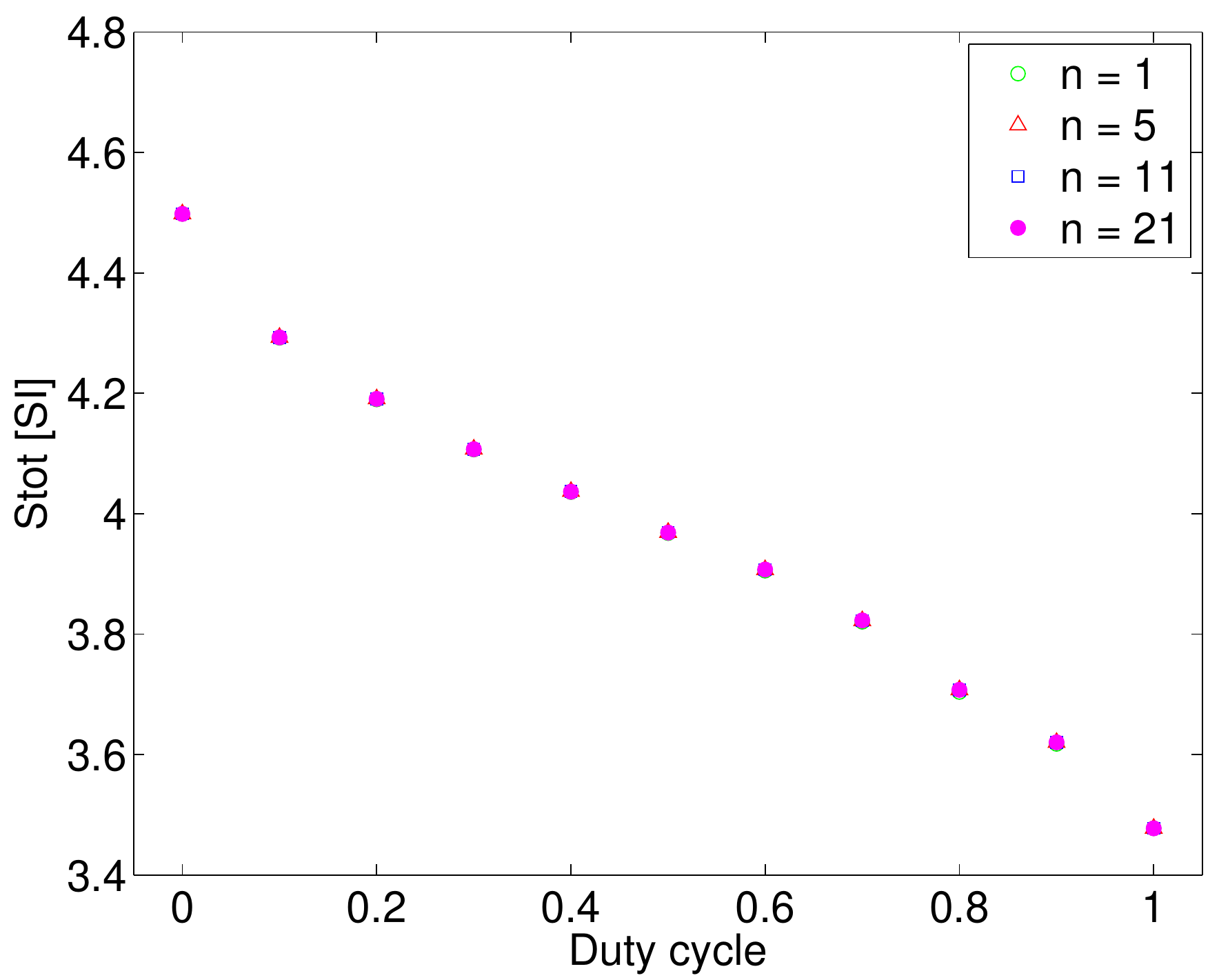}

\label{newd_10um_Per_0.1um_fig}}\\

\par\end{centering}

\caption{Spectral contributions to the thermal capacitance and total thermal
capacitance for the structure shown in Fig. \ref{rods_fig} with d
= 10$\mu$m and different values of duty cycle with the periodicity
of (a) Per = 10$\mu$m (b) Per = 1$\mu$m (c) Per = 0.1$\mu$m}
\label{newd_10um_fig}
\end{figure*}

\begin{figure*}
\begin{centering}
\subfloat[]{\includegraphics[scale=0.35]{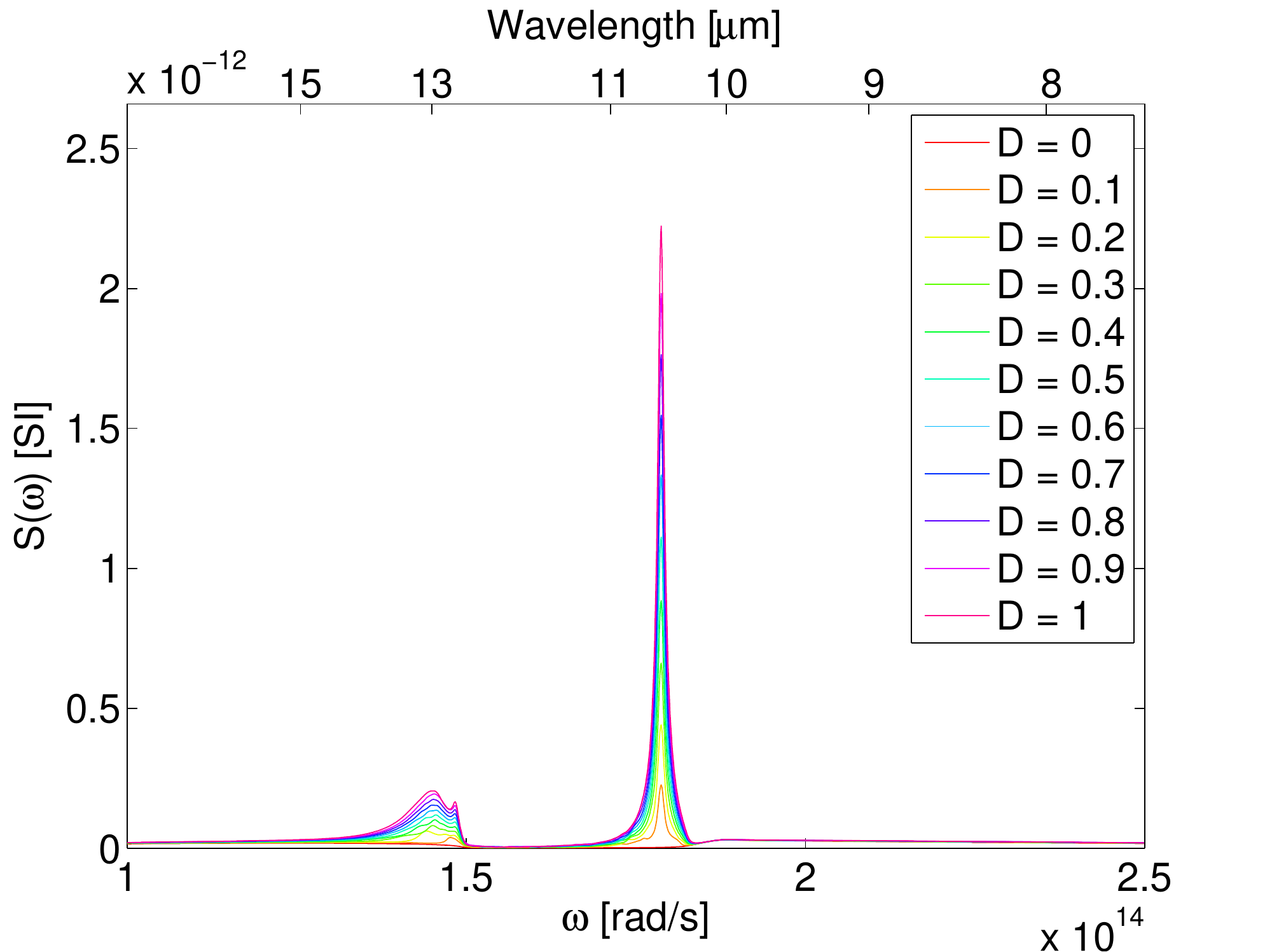}\includegraphics[scale=0.35]{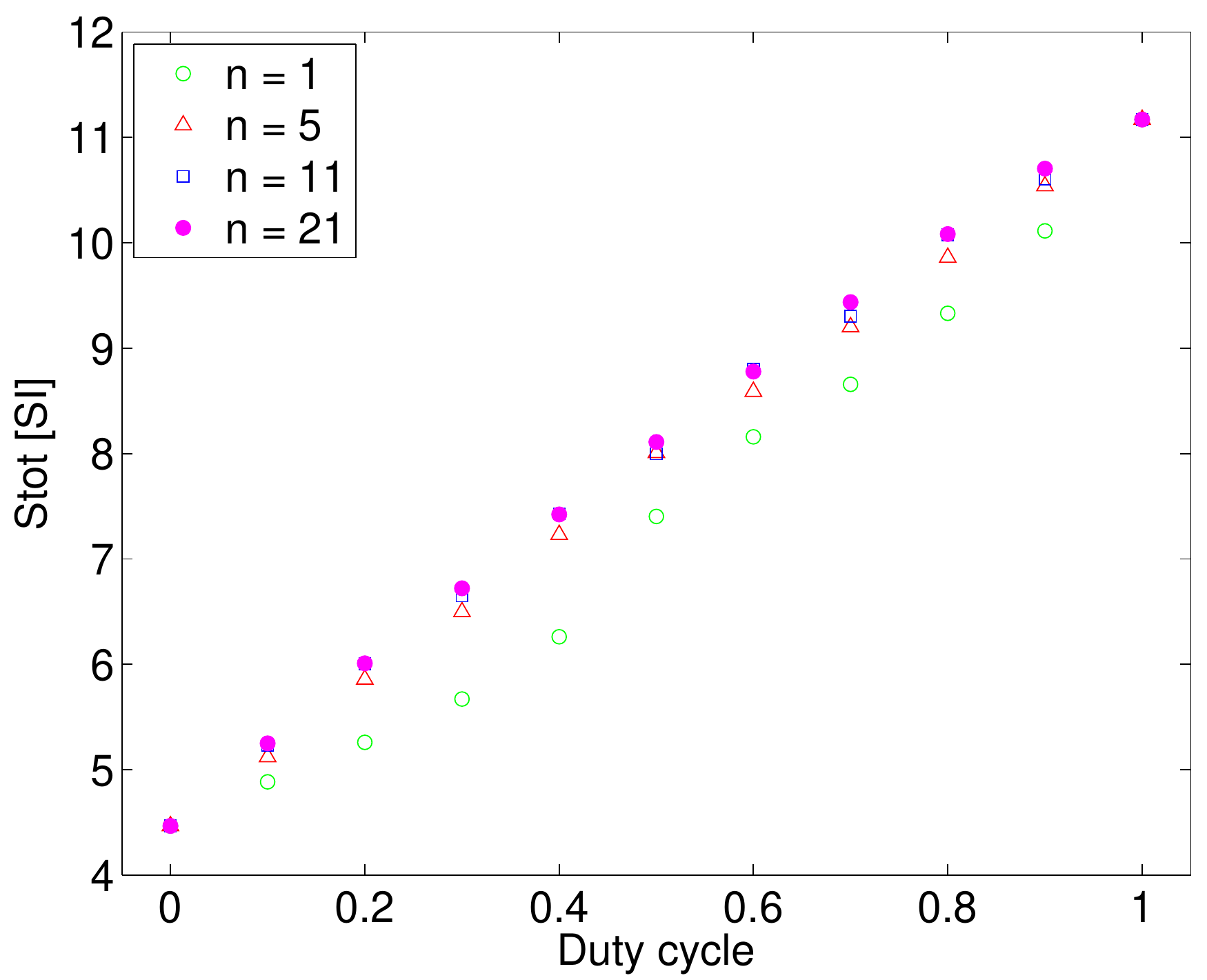}\label{newd_1um_Per_10um_fig}}\\
\subfloat[]{\includegraphics[scale=0.35]{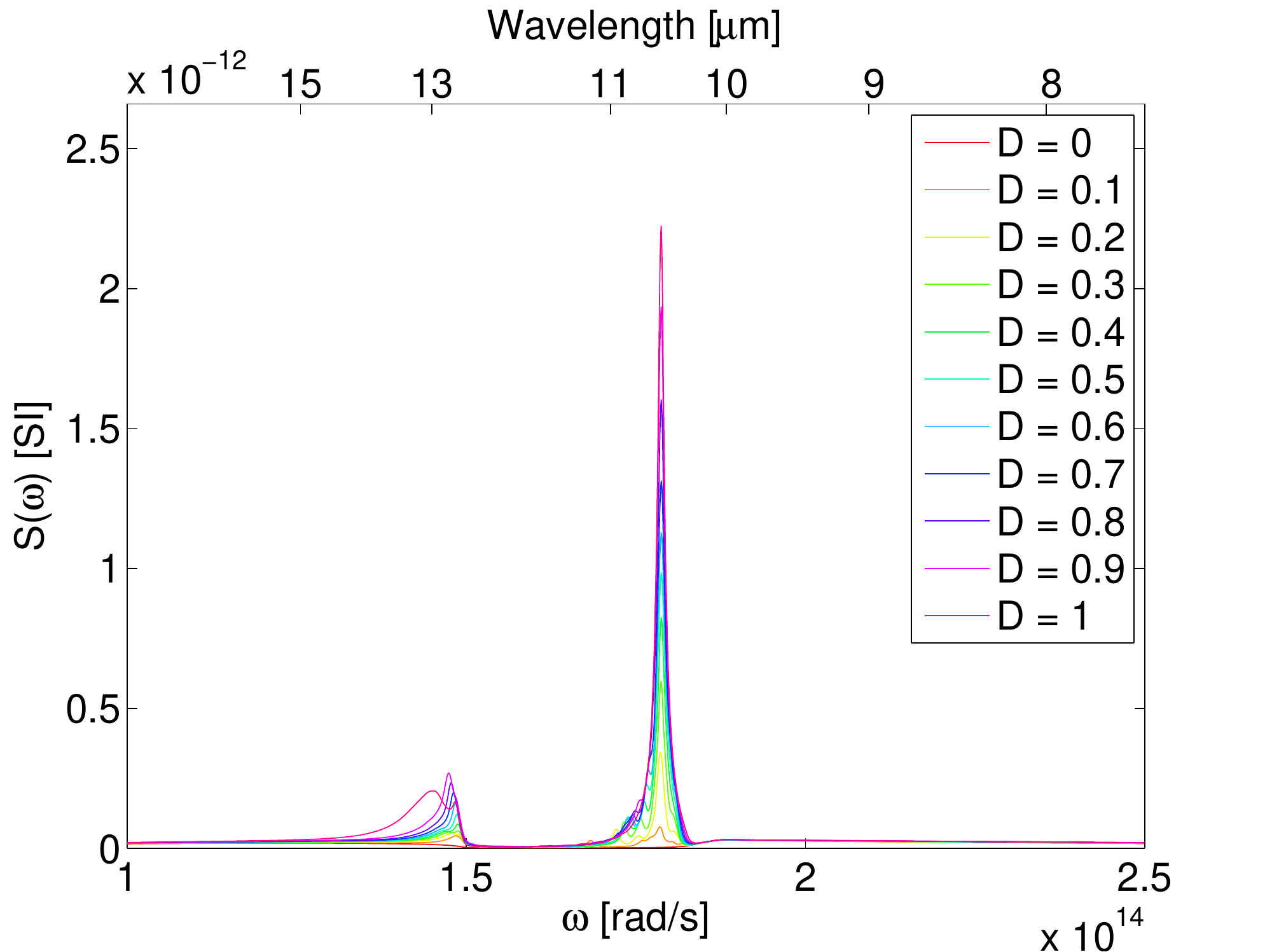}\includegraphics[scale=0.35]{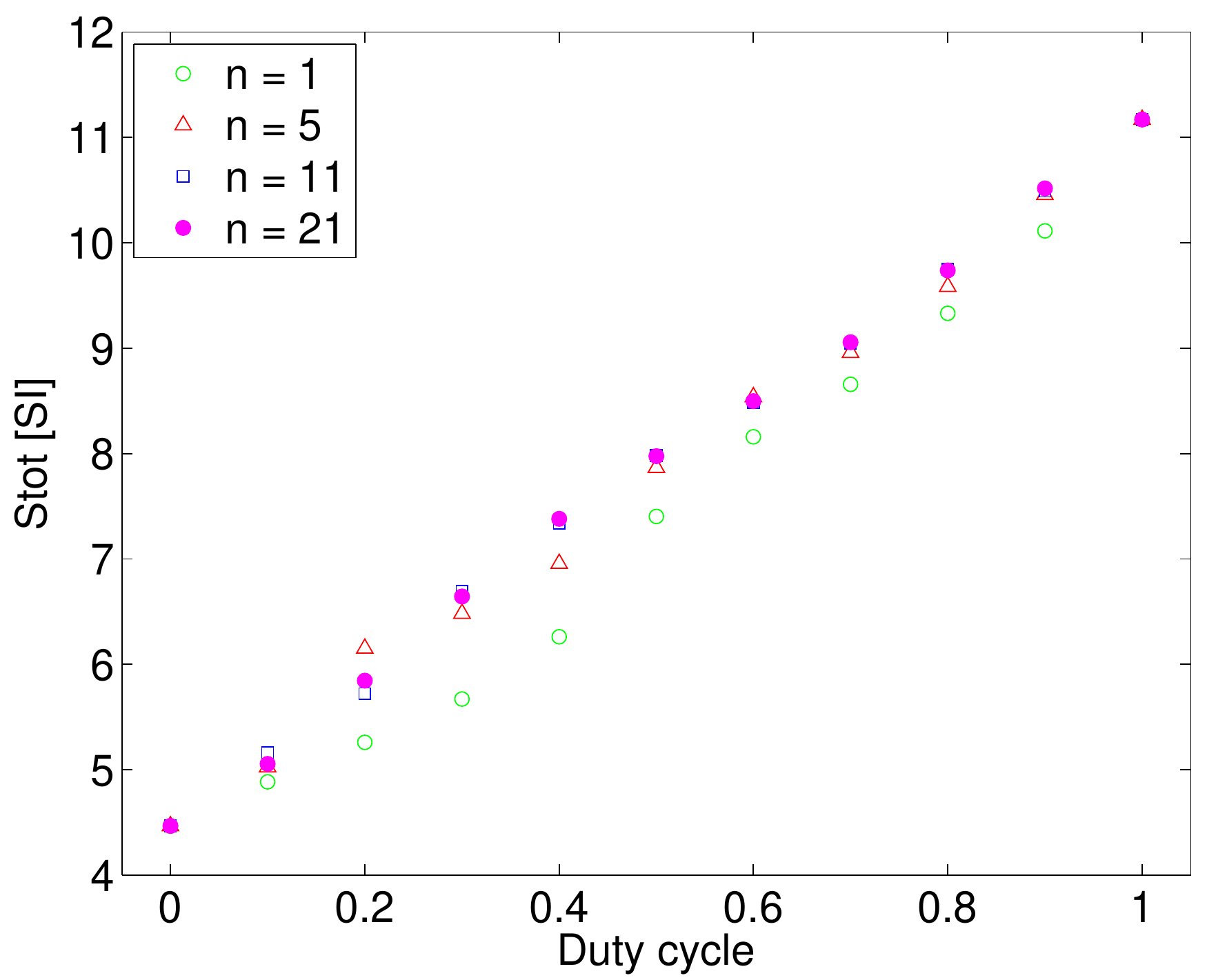}

\label{newd_1um_Per_1um_fig}}\\
\subfloat[]{\includegraphics[scale=0.35]{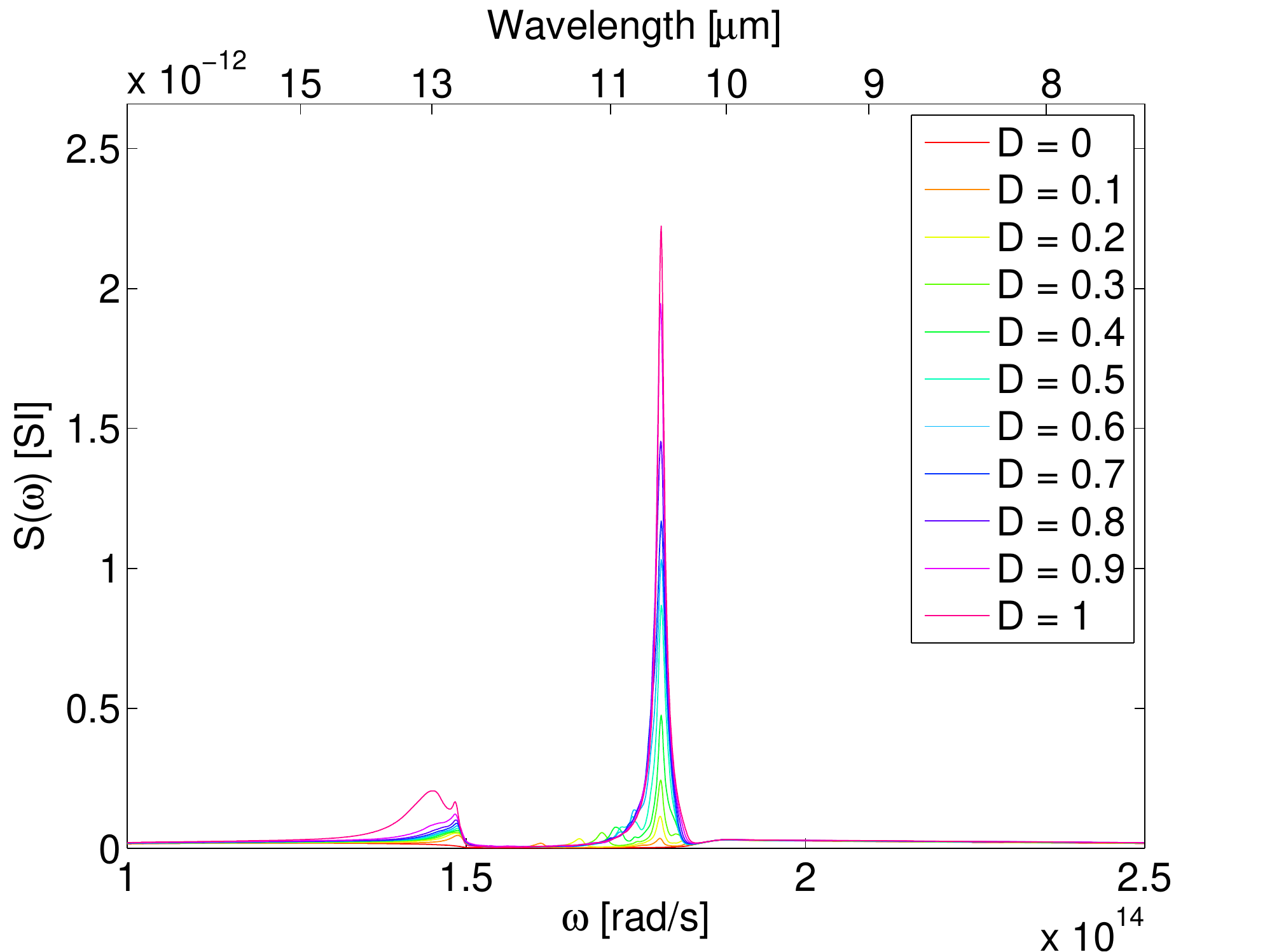}\includegraphics[scale=0.35]{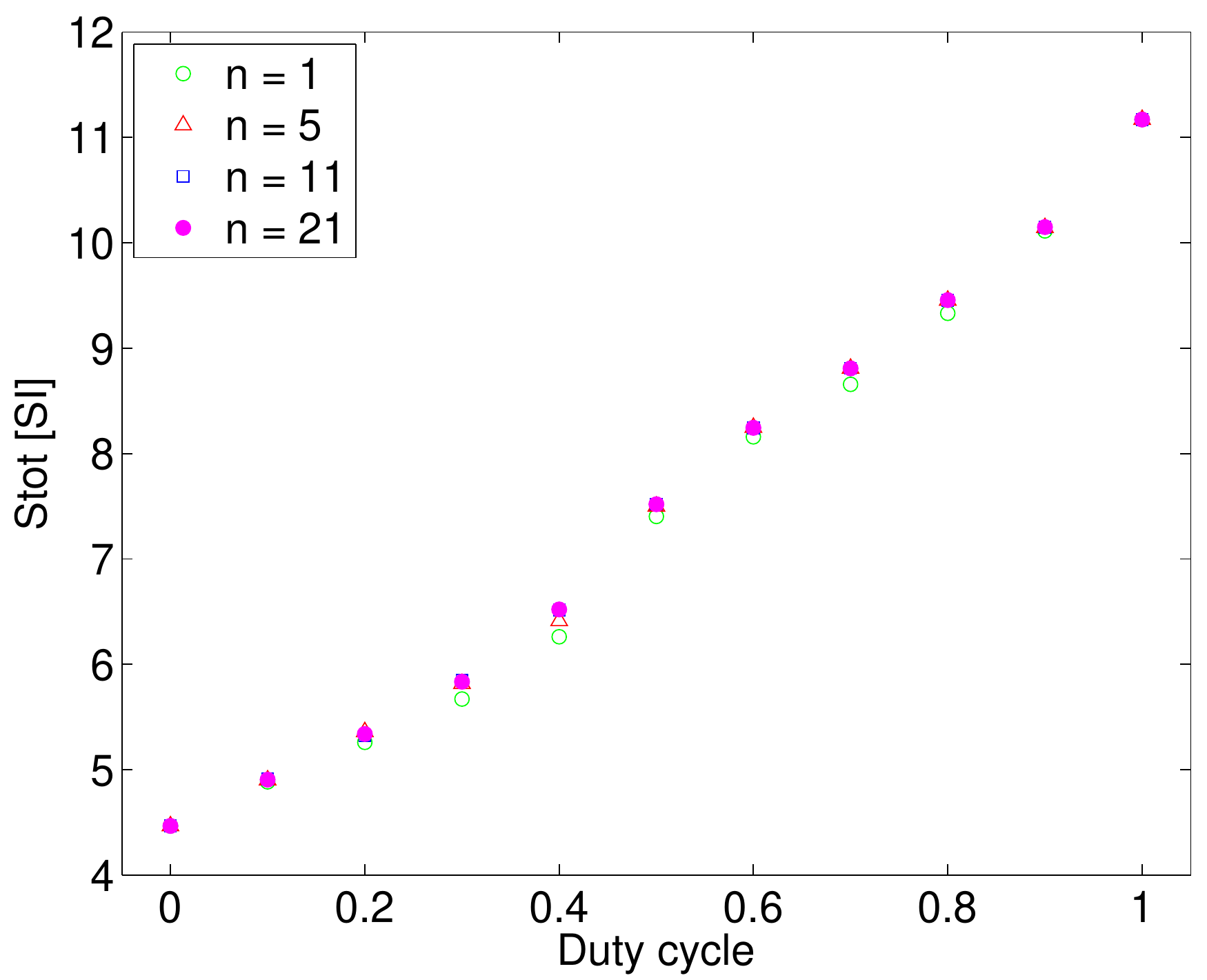}

\label{newd_1um_Per_0.1um_fig}}\\

\par\end{centering}

\caption{Spectral contributions to the thermal capacitance and total thermal
capacitance for the structure shown in Fig. \ref{rods_fig} with d
= 1$\mu$m and different values of duty cycle with the periodicity
of (a) Per = 10$\mu$m (b) Per = 1$\mu$m (c) Per = 0.1$\mu$m}
\label{newd_1um_fig}
\end{figure*}

\begin{figure*}
\begin{centering}
\subfloat[]{\includegraphics[scale=0.35]{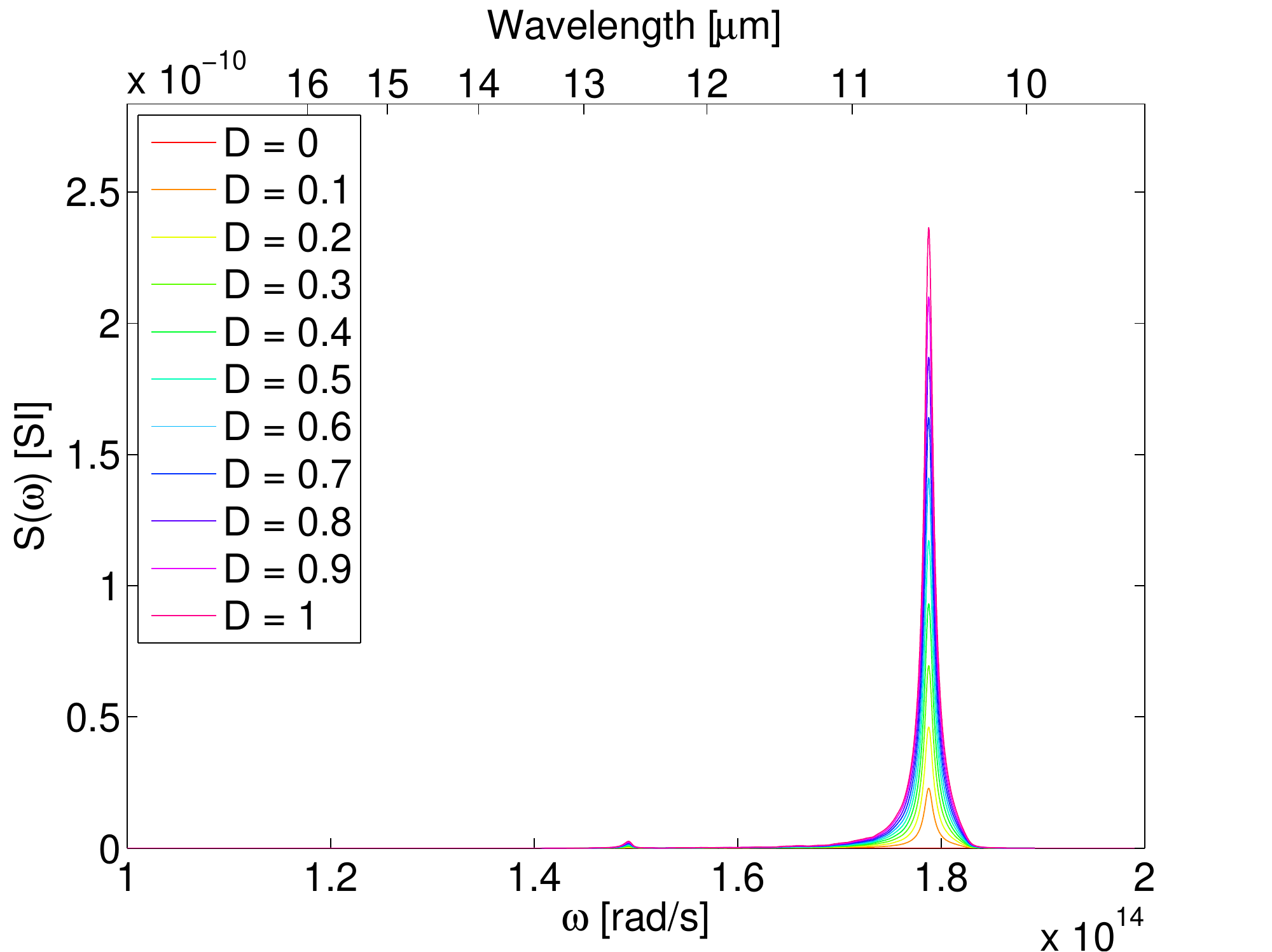}\includegraphics[scale=0.35]{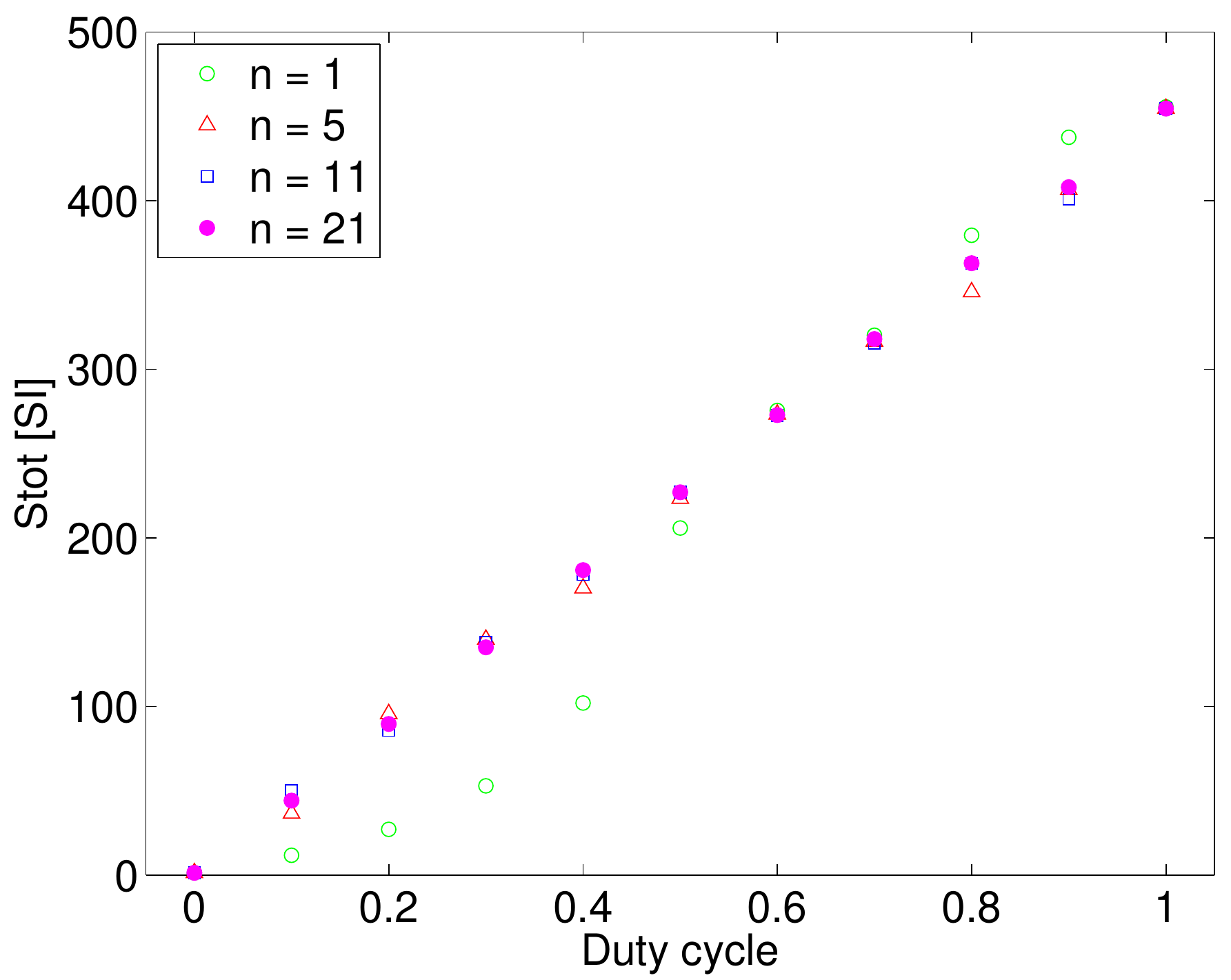}\label{newd_0.1um_Per_10um_fig}}\\
\subfloat[]{\includegraphics[scale=0.35]{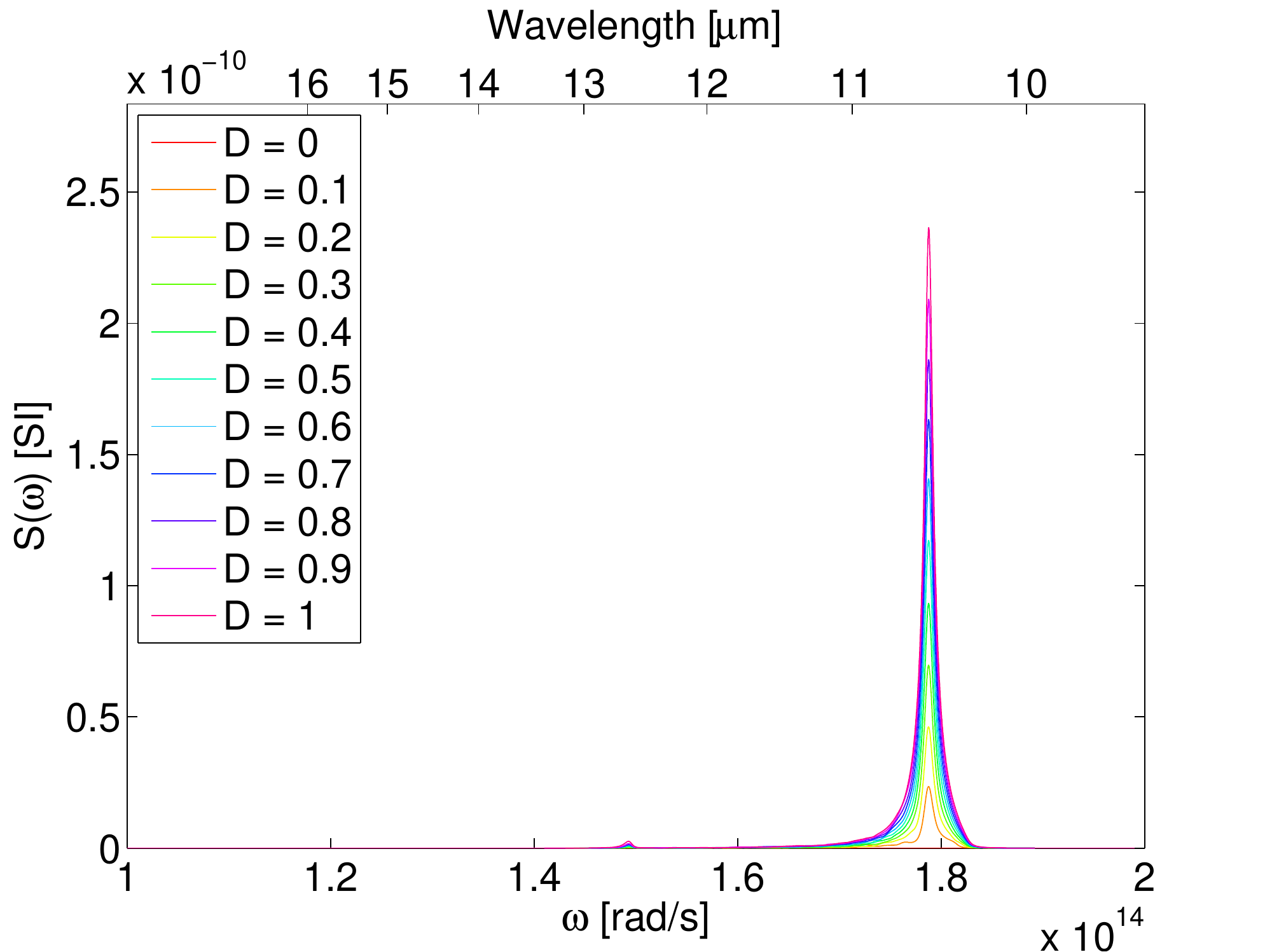}\includegraphics[scale=0.35]{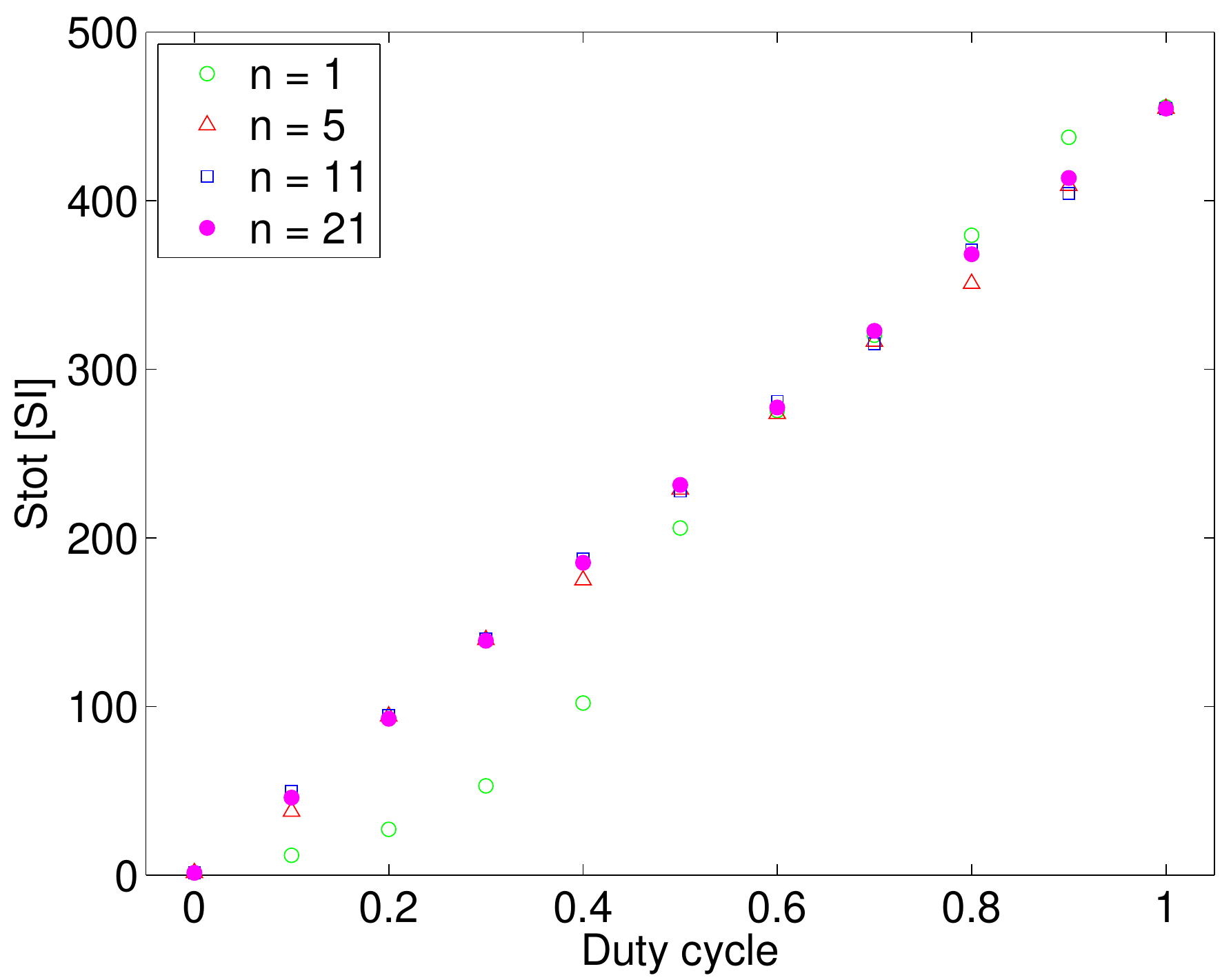}

\label{newd_0.1um_Per_1um_fig}}\\
\subfloat[]{\includegraphics[scale=0.35]{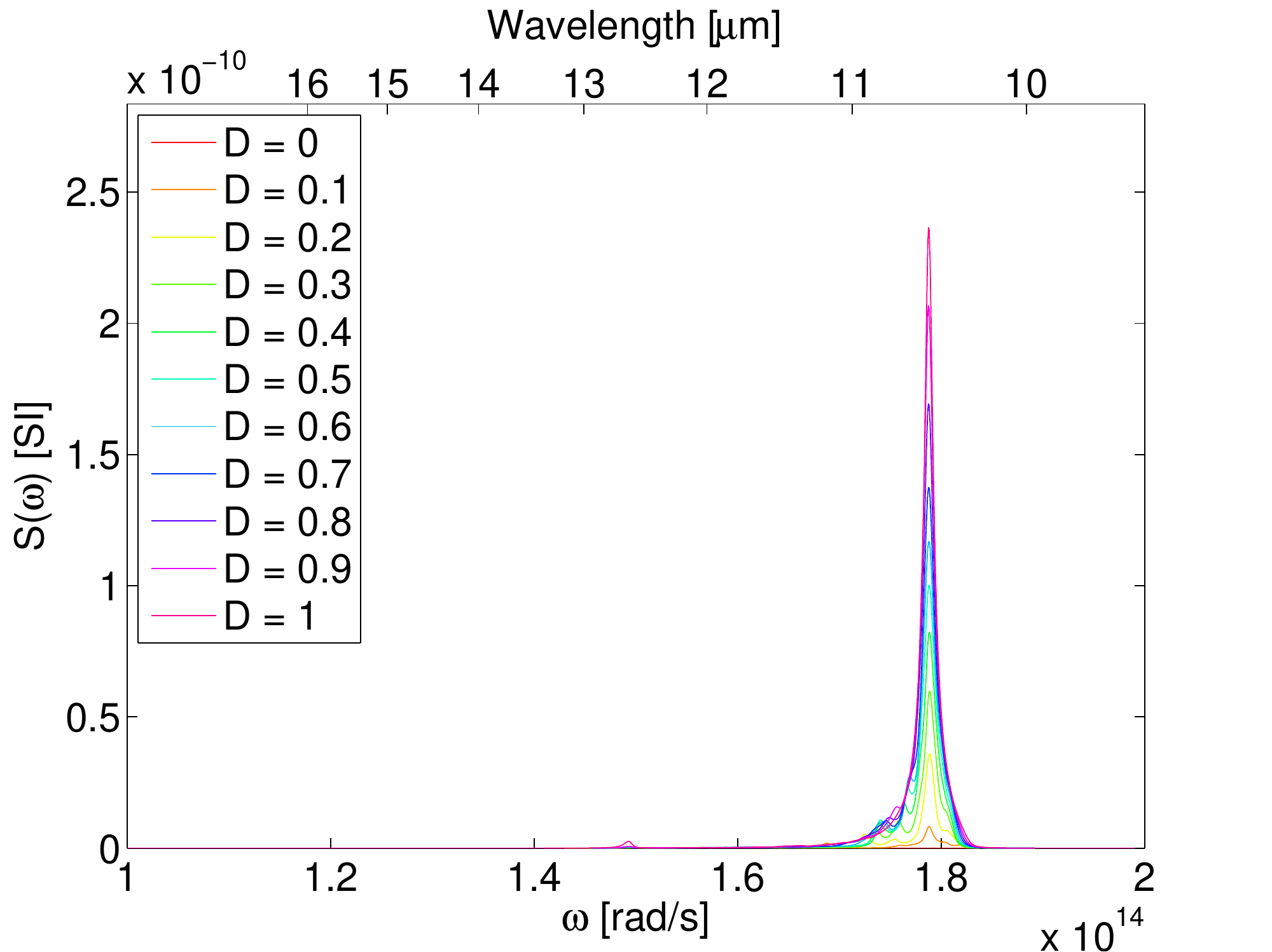}\includegraphics[scale=0.35]{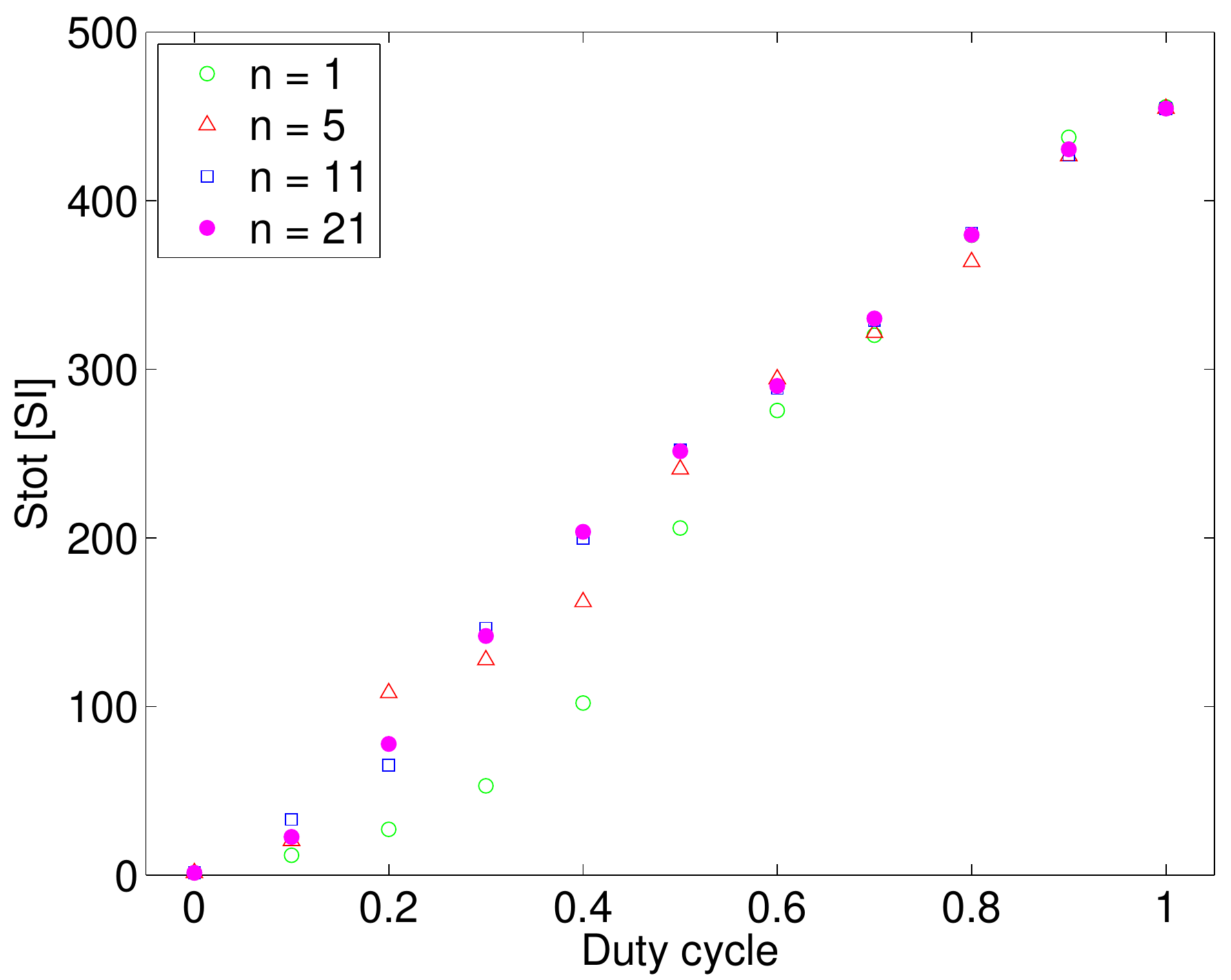}

\label{newd_0.1um_Per_0.1um_fig}}\\

\par\end{centering}

\caption{Spectral contributions to the thermal capacitance and total thermal
capacitance for the structure shown in Fig. \ref{rods_fig} with d
= 0.1$\mu$m and different values of duty cycle with the periodicity
of (a) Per = 10$\mu$m (b) Per = 1$\mu$m (c) Per = 0.1$\mu$m}
\label{newd_0.1um_fig}
\end{figure*}

In the second structure we have considered, the thermal capacitance
of the SiC beams, and a slab of SiC is numerically calculated. The
details of the notations used for the parameters involved in this
structure are shown in Fig. \ref{rods_fig}. Results of the thermal
capacitance corresponding to $d=10\mu m$, $d=1\mu m$, and $d=0.1\mu m$
are shown in Figs \ref{newd_10um_fig}, \ref{newd_1um_fig}, and \ref{newd_0.1um_fig},
respectively. In the structures considered, $d_{rod}=0.5d$, is assumed.
Like the previous structure, the thermal capacitance across a period
is plotted for different distances and periodicities, using D = 0.4
in all cases. The plots which are shown in Fig. \ref{newd_xvar_fig}
were all obtained by using 21 harmonics.

One important fact regarding this structure is that we should obtain
the thermal emission from a slab of SiC to the vacuum for the case
of a duty cycle of 0. (In this case there are no beams anymore) Our
formalism nicely reproduces this result in this case. Again like for
the previous structure we see the monotonically increase in thermal
capacitance for the distances of $d=1\mu m$, and $d=0.1\mu m$. Note
that this increase is not necessarily linear with the duty cycle.
However this increase becomes more linear for large values of periodicity.
This again is consistent with our intuition that for large values
of periodicity, the cross talk between neighboring beams are negligible. 

The situation for $d=10\mu m$ with $Per=10\mu m$ is more interesting.
In this case as we encountered previously for the grating structure,
the extremum of thermal capacitance is achieved for a duty cycle which
is neither zero or unity. Here, we again expect the Mie resonances
of the nanobeams come into play. Note that in this case, the result
of effective medium theory, has the largest inaccuracy. This is expected
since in this case the periodicity is the largest compared with the
two other cases for $d=10\mu m$. (Specifically, cases of $Per=1\mu m$
and $Per=0.1\mu m$)

One important fact about our method is that it can be used in this
way for calculation of thermal transfer between a slab and a particle
with an arbitrary shaped structure. This comes from the fact that
when the periodicity becomes large, the cross talk between particles
becomes negligible and the thermal capacitance is coming from the
sum of the contributions of individual beams. This can be proposed
as an alternative method for calculation of thermal capacitance between
e.g. a sphere and a slab that has been done in several methods in
several references \citep{PhysRevLett.106.210404,PhysRevB.84.245431,PhysRevB.85.165104}.

\section*{Conclusions:}

In this paper, we have developed a formalism for calculating the thermal
transfer of periodic structures with building blocks of arbitrary
size and shape. We applied this method to obtain the thermal capacitance
between a slab of SiC and binary SiC gratings. We also used this method
for the calculation of the thermal transfer between a plain slab of
SiC and an array of SiC beams of rectangular cross section. The obtained
results show that, thermal capacitance in these cases can accurately
be obtained through incorporation of some of the first harmonics.
Moreover, results show that the thermal transfer changes monotonically
with increasing duty cycle for the cases that distances are much smaller
than the resonance wavelength. However, this trend breaks in the case
that distances are on the same order of magnitude as the resonance
wavelength. 

Our method, in the case of incorporating just one harmonic reproduces
the results obtained by the effective medium theory. In this regard,
this method can be used to determine the accuracy of the effective
medium theory for specific structures of interest. According to the
numerical results obtained, as we expect, by decreasing the periodicity
of the structure to the subwavelength regime compared with the relevant
resonance wavelengths in the system, effective medium theory becomes
increasingly accurate.

This method can also be used to analyze the thermal transfer between
structures in which one of the materials is composed of an array of
particles. Since in the limit of large periodicity, the cross talk
between particles becomes negligible, this method poses itself to
be used for calculation of thermal transfer between a slab and arbitrary
shaped particles. For the reasons above, we believe that the presented
technique will prove versatile for calculating and optimizing the
thermal transfer between a wide variety of practical structures.

\bibliographystyle{apsrev4-1}
\bibliography{allbibs_wlinks}

\end{document}